\documentclass[10pt]{iopart}

\usepackage{graphicx}   
\usepackage{color}      
\usepackage{upgreek}
\usepackage{cite}
\usepackage[percent]{overpic}

\newcommand{\Wkm}{W$^{-1}$~km$^{-1}$ }
\newcommand{\WkmN}{W$^{-1}$~km$^{-1}$}

\newcommand{\mum}{$\upmu$m }
\newcommand{\mumN}{$\upmu$m}

\newcommand{\pskm}{ps$^2$~km$^{-1}$ }
\newcommand{\pskmN}{ps$^2$~km$^{-1}$}

\usepackage[overlay,absolute]{textpos}
\newcommand\PlaceText[3]{%
	\begin{textblock*}{10in}(#1,#2)
		#3
	\end{textblock*}
}%

\begin{document}
\title{Dispersion engineering of mode-locked fibre lasers [Invited]}

\author{R. I. Woodward}
\address{MQ Photonics, Department of Engineering, Macquarie University, New South Wales, Australia}
\ead{robert.woodward@mq.edu.au}

\vspace{10pt}

\begin{abstract}
Mode-locked fibre lasers are important sources of ultrashort pulses, where stable pulse generation is achieved through a balance of periodic amplitude and phase evolutions.
A range of distinct cavity pulse dynamics have been revealed, arising from interplay between dispersion and nonlinearity in addition to dissipative processes such as filtering.
This has led to the discovery of numerous novel operating regimes, offering significantly improved laser performance.
In this Topical Review, we summarise the main steady-state pulse dynamics reported to date through cavity dispersion engineering, including average solitons, dispersion-managed solitons, dissipative solitons, giant-chirped pulses and similaritons.
Characteristic features and the stabilisation mechanism of each regime are described, supported by numerical modelling, in addition to the typical performance and limitations.
Opportunities for further pulse energy scaling are discussed, in addition to considering other recent advances including automated self-tuning cavities and fluoride-fibre-based mid-infrared mode-locked lasers.
\end{abstract}

\PlaceText{20mm}{20mm}{Journal of Optics, Vol. 20, 033002 (2018); https://doi.org/10.1088/2040-8986/aaa9f5}

\vspace{2pc}
\noindent{\it Keywords}: fibre lasers, mode-locked lasers, ultrafast lasers, pulse shaping, solitons
%
\ioptwocol

\section{Introduction}

The generation of ultrashort optical pulses is an enabling technology that has opened up many applications across science, industry and medicine.
Mode-locked lasers, which generate pulses of coherent light through phase-locking of cavity modes, are at the forefront of this field.
Despite great progress over the past half-century, however, end-users requirements for these sources still continue to grow as the breadth of applications increases, driving sustained research in this area.

Today, there is particular demand for high-power mode-locked \emph{fibre} lasers, where rare-earth-doped fibres offer a number of advantages over alternative gain media, including excellent beam quality, simple thermal management and a compact and robust design~\cite{Okhotnikov}.
To achieve high powers, a master-oscillator-power-fibre-amplifier (MOPFA) architecture is typically employed, comprising a  mode-locked fibre oscillator followed by multiple amplifier stages~\cite{Fermann2003}.
The performance, efficiency and simplicity of such pulse sources could be improved, however, by augmenting the pulse energy generated directly by the oscillator, minimising the need for subsequent amplification.
Intense research activity is thus considering new mode-locked cavity designs, pulse shaping dynamics and associated nonlinear limitations.

Dispersion (and its interplay with nonlinearity) is a principal factor influencing the pulse dynamics in mode-locked fibre lasers.
While early studies (e.g. using dye lasers~\cite{DeSilvestri1984}) established an understanding of the role of cavity dispersion on ultrashort-pulse shaping, the flexibility of the guided-wave fibre geometry and unprecedented ability to engineer cavity dispersion through choice and length of fibre components, has resulted in the discovery of numerous new pulse behaviours and laser operating regimes.
In addition to delivering record performance, these studies have contributed to basic science through unveiling new understanding of the underlying nonlinear wave dynamics.

In this Topical Review, we summarise the current state-of-the-art in dispersion engineering of mode-locked fibre lasers with a particular focus on various operating regimes that have emerged.
A brief overview of the relevant theoretical concepts of pulse formation, shaping and de-stabilisation mechanisms is first presented in Section~\ref{sec:theory}.
Understanding of these phenomena and the ability to control nonlinearity using dispersion has unveiled various parameter spaces that achieve distinct pulse solutions, including solitons, dispersion-managed solitons, dissipative solitons, giant-chirped pulses and similaritons. 
Section~\ref{sec:designs} categorises the major designs and operating regimes to date, highlighting leading reported performances and elucidating the cavity dynamics through numerical modelling.
Finally, we discuss outstanding challenges in the field and comment on emerging trends and opportunities in Section~\ref{sec:outlook}.

\section{Theory of Pulse Shaping in Mode-Locked Fibre Lasers}
\label{sec:theory}

\subsection{Steady-State Pulse Formation}
Laser oscillators are complex nonlinear dissipative systems, which can support steady-state solutions in the form of stable optical pulses---providing both amplitude and phase changes from interactions with cavity components are balanced.
Within each cavity round trip, a pulse can experience saturable gain / loss, spectral filtering, dispersive spreading and a range of nonlinear phenomena related to the Kerr effect and stimulated scattering.

\begin{figure}[b]
	\centering
	\includegraphics{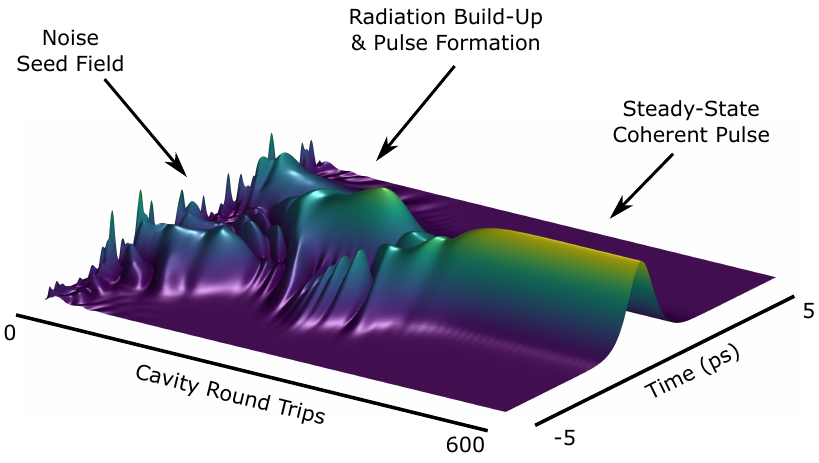}
	\caption{Simulated pulse build-up in a mode-locked fibre laser from noise, over hundreds of cavity round trips.}
	\label{fig:evol_from_noise}
\end{figure}

Stable pulse generation relies on careful laser design to achieve compensation between the various effects and thus, a self-consistent pulse evolution within the cavity that repeats each round trip, building up and stabilising from an initial noise field (Fig.~\ref{fig:evol_from_noise}).
Practically, mode-locking is initiated by periodically modulating the amplitude or phase of light at the fundamental cavity free-spectral range; this can be implemented actively using an electronically controlled modulator, or passively using a saturable absorber device with a nonlinear absorption profile.
The nonlinear switching speed can also play an important role in the steady-state cavity dynamics~\cite{Kartner1998}, although `fast' saturable absorbers (e.g. using semiconductors, nanomaterials or fibre-based nonlinear effects~\cite{Keller1996,Martinez2013a,Woodward2015_as_2d,Sobon2015}) are now commonplace, where the modulator response time has a minimal influence on pulse shaping and the steady-state properties, compared to factors such as dispersion and nonlinearity. 

\subsection{Dispersion \& Nonlinearity}
While the term `dispersion' encompasses a range of linear propagation effects arising from the frequency dependence of a guided mode's effective refractive index, including polarization-mode dispersion, intermodal dispersion and chromatic dispersion~\cite{Agrawal2013}, it is the latter phenomena that is of most importance for typical mode-locked lasers using single-mode fibre, particularly the second-order term which describes group-velocity dispersion (GVD) $\beta_2$.
While higher-order terms can become important as the GVD approaches zero~\cite{Dennis1994a}, the main pulse shaping regimes and dynamics are primarily determined by the sign and magnitude of GVD; thus, higher-order dispersion is not considered further herein.
Transform-limited pulse propagation in the presence of GVD alone broadens the pulse duration, with accumulation of quadratic phase and thus, a linear frequency sweep (chirp) developed across the pulse. 
By compensating this chirp with the opposite sign of GVD, the pulse can be recompressed back to the transform limit.

From a laser design point of view, there are various options for dispersion managing a cavity.
Standard step-index silica fibre typically has a zero GVD wavelength $\sim$1.27~\mum from the material dispersion contribution, although waveguide dispersion tailoring by varying the fibre refractive index profile / using novel microstructured designs can enable a wide range of dispersive properties.
Similarly, chirped fibre Bragg gratings can be used to introduce a specific amount of dispersion.
In addition to fibre-based options, diffraction gratings and prism pair dispersive delay lines can be included, albeit at the cost of a fully-fibre integrated design.

The long interaction length of high-intensity pulses in fibre also results in strong nonlinear phenomena arising from the Kerr effect.
Fibres are characterised by an effective nonlinear parameter $\gamma(\lambda) = 2 \pi n_2 / (\lambda  A_\mathrm{eff})$ where $A_\mathrm{eff}$ is the effective mode area and $n_2$ is the material nonlinear index.
At the simplest level, nonlinear phase accumulation during pulse propagation can lead to spectral broadening through self-phase modulation (SPM).
The resulting chirp profile is nonlinear, however, which is more challenging to compensate.
Additional nonlinear effects can manifest from stimulated Raman scattering (Section~\ref{sec:nl_instab}).

Dispersion and nonlinearity cannot be considered independently, however, and in mode-locked fibre lasers the interplay between these effects gives rises to particularly interesting and useful phenomena, such as solitons.

\subsubsection{Solitons}
\label{sec:solitons}
If the dispersive and nonlinear phase shifts exactly cancel out, a pulse is able to propagate over long distances while preserving both its temporal and spectral shape.
Such a solitary wave is known as a \emph{soliton}, which is an analytical solution of the Nonlinear Schr\"{o}dinger Equation (NLSE) for the case of anomalously dispersive (negative $\beta_2$) fibre~\cite{Zakharov1972, Hasegawa1973}.
Solitons are unchirped and exhibit a characteristic hyperbolic secant amplitude profile:
\begin{equation}
A(t) = \sqrt{P_0} ~ \mathrm{sech}\left(\frac{t}{\tau}\right)
\end{equation}
where $P_0$ is the peak power and $\tau$ is the soliton width ($\mathrm{FWHM~duration}=1.76\tau$).
We should also carefully clarify the phase cancellation mechanism here: large values of dispersion always result in a linear chirp independent of pulse shape, although the profile of the dispersive phase does vary with pulse shape and propagation distance for small values.
While Gaussian-shaped pulse dispersive chirp is always linear, the chirp accumulation of a sech pulse over a short distance is actually nonlinear, which explains how anomalous dispersion is able to exactly cancel out the nonlinear chirp from SPM to form a soliton~\cite{Dudley2001}.

The pulse energy of a soliton is firmly defined by an `area theorem' which relates energy to the soliton width and fibre properties~\cite{Zakharov1972, Hasegawa1973}:
\begin{equation}
\label{eqn:area_theorem}
E = \frac{2|\beta_2|}{\gamma\tau}.
\end{equation}

Thus, for a given fibre and pulse duration, the soliton energy is fixed. 
Pulses with lower energies will not form solitons and will broaden dispersively.
Pulse energies exceeding this value will result in shedding of energy (into a `dispersive wave') to achieve the required pulse energy for a fundamental soliton, or for pulse energies above the square of an integer number times the energy in Eqn.~\ref{eqn:area_theorem}, the formation of a higher-order soliton.
The shape of higher-order solitons varies periodically as they propagate, although disturbances (gain/loss, higher-order dispersion etc.) cause them to split into multiple fundamental solitons~\cite{Taylor1992}.

Fundamental solitons are also stable against changes in their environment: they adiabatically tailor their shape during propagation providing the length scale of the change is long compared to the soliton period, defined as the distance for a $\pi/4$ phase delay, $Z_0=\pi \tau^2/(2\beta_2)$.
Additionally, a propagating soliton which is periodically disturbed (e.g. circulating in a laser cavity) can be considered to experience the path-averaged value for dispersion and nonlinearity---referred to as an `average soliton'~\cite{Kelly1991}, which is of critical importance in mode-locked soliton laser dynamics (Section~\ref{sec:desgin_sol}).

As new laser designs with distinct dispersion maps, pulse shaping mechanisms and intracavity dynamics have emerged, researchers have sought to categorise them, leading to the modern nomenclature for describing types of pulses.
Therefore, before describing these phenomena in Section~\ref{sec:designs}, we clarify the term soliton in this context. 
From a strictly mathematical definition, solitons are solutions of completely integrable nonlinear partial differential equations (e.g. the NLSE, which only includes GVD and SPM)~\cite{Zakharov1972}.
Pulse propagation along low-loss fibre where GVD and SPM are the dominant effects can be considered as a conservative Hamiltonian system, therefore possessing soliton solutions.
A mode-locked resonator, however, is a dissipative system exhibiting periodic gain and loss.
Thus, the pulses formed from such a laser are not, in the mathematical sense, solitons.
Practically, however, the optics community often refers to any localised, persistent structure that is robust to a range of perturbations as a soliton~\cite{Kutz2006}.

\subsection{Instabilities and Limiting Factors}
As pulse energies in a laser cavity are increased, nonlinear instabilities can break the steady-state amplitude/phase balance, manifesting in the fission of single coherent pulses to form chaotic multiple pulsing emission, or decoherence of phase-locked cavity modes leading to a loss of mode-locking.
An intermediate region of noise burst (or `noise-like pulse') operation  can also exist, which can be interpreted as partial mode-locking~\cite{Bradley1974}, where pulses with partially linear (or occasionally completely random~\cite{Ozgoren2010}) chirp are generated~\cite{Horowitz1997}; while such pulses can be useful for certain applications~\cite{Ozgoren2010}, they are noisy, not compressible~\cite{Jeong2014a} and thus for the scope of this review, considered a negative phenomena.
We now briefly outline the major limiting effects in mode-locked fibre lasers.

\subsubsection{Soliton Limitations}
\label{sec:nl_instab_sol}

Long-distance soliton communications work with periodic amplification demonstrated that solitons exhibit a resonant instability when the disturbance length scale $L$ (i.e. amplifier spacing) approaches 8$Z_0$, due to phase matching between the soliton and dispersive wave it sheds when perturbed~\cite{Mollenauer1986}.
A mode-locked fibre laser with amplification each cavity round trip length $L$ exhibits the same limitation, resonantly transferring energy out of the pulse to dispersive radiation (leading to spectral sideband formation) unless $L<<8Z_0$~\cite{Kelly1992}.
This limits the shortest possible pulse width as $Z_0\propto\tau^2$.

Additionally, for any given fibre, $\beta_2$ and $\gamma$ are fixed, thus the relationship between soliton duration and energy is set by the area theorem (Eqn.~\ref{eqn:area_theorem}), as illustrated in Fig.~\ref{fig:instabilities}a for standard telecommunications fibre.
This prevents scaling of pulse energies as excess energy will be shed.

\subsubsection{Wave Breaking}
\begin{figure}[tb]
	\centering
	\sffamily
	\begin{overpic}{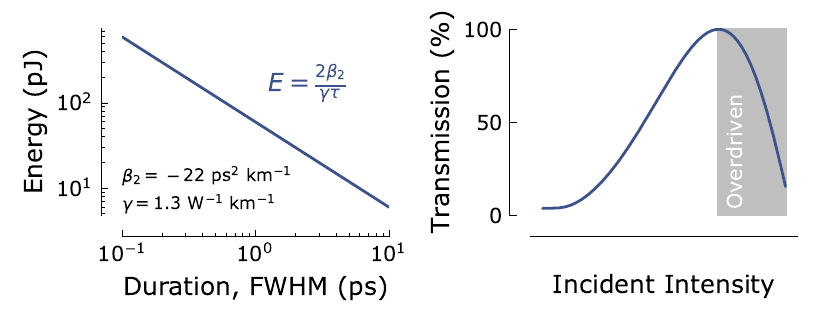}
		\put(0, 36){ {\small (a)} }
		\put(44, 36){ {\small (b)} }
		\put(0, 0){ {\small (c)} }
	\end{overpic}
	\rmfamily
	\includegraphics{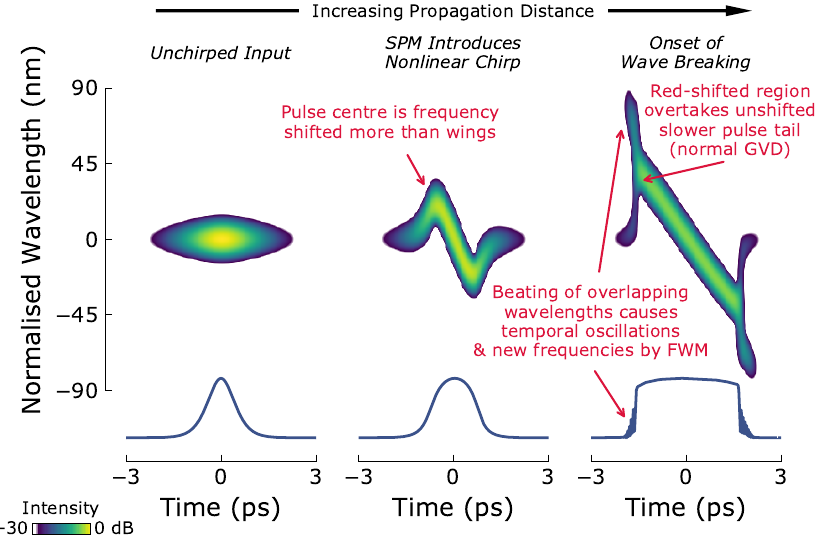}
	\caption{Instabilities and limiting factors: (a) soliton energy--duration relationship for typical telecommunications fibre; (b) overdriving of an artificial saturable absorber; (c) simulation of wave breaking during pulse propagation.}
	\label{fig:instabilities}
\end{figure}

To prevent conventional soliton formation, anomalously dispersive fibre can be avoided.
However, interplay between normal GVD (positive $\beta_2$) and SPM can also lead to a destabilising phenomena for high-energy pulses known as `wave breaking'~\cite{Tomlinson1985, Anderson1992}.
Fig.~\ref{fig:instabilities}c illustrates this behaviour through simulated spectrograms.
Firstly, nonlinear fibre propagation creates a nonlinear chirp (i.e. non-monotonic frequency sweep) through SPM: light near the leading (trailing) edge of the pulse is red (blue) shifted, while the low-intensity pulse wings are not frequency shifted.
The red (blue) shifted pulse regions thus travel faster (slower) than the wings due to normal dispersion, leading them to temporally overlap.
The resulting interference induces temporal oscillations and the generation of new frequencies in the spectrum by four-wave mixing (FWM).
In terms of cavity dynamics in a mode-locked laser, this wave breaking phenomena can destabilise the pulse and phase balance, leading to a loss of coherent mode-locking.

\subsubsection{Overdriving of Saturable Absorbers}
Saturable absorbers exhibit a nonlinear transmission curve with increasing transmittance for greater incident peak intensities, preferentially promoting the generation of high-power pulses over low-intensity CW light.
For \emph{artificial} saturable absorbers based on fibre nonlinearity, such as nonlinear polarization evolution (NPE)~\cite{Stolen1982b} and nonlinear optical/amplifying loop mirrors (NOLM/NALMs)~\cite{Doran1988, Fermann1990}, where the transmission function is produced interferometrically, the transmission curve becomes periodic at high incident intensities.
Beyond a certain point, therefore, the gradient becomes negative and the saturable absorber is said to be `overdriven' (Fig.~\ref{fig:instabilities}b).

This effect can destabilise mode-locking or manifest as a peak-power clamping effect, whereby it is energetically unfavourable for the cavity to permit higher peak powers (as this would decrease absorber transmission), thus any increase in pulse energy is accommodated by pulse broadening, often with random chirp so the pulse is incompressible.
The saturable absorber could be biased to operate at higher powers but this may inhibit self-starting from low-intensity noise fields. 
It should also be noted that reverse saturable absorption at high intensities has been observed in \emph{real} saturable absorbers such as semiconductor saturable absorber mirrors and nanomaterials through two-photon absorption (TPA)~\cite{Jiang1999, Woodward_prj_2015}.

In terms of saturable absorbers, it has also been shown numerically~\cite{Renninger2015} and qualitatively in the framework of the Haus Master Equation~\cite{Haus2000} that the modulation depth of the absorber directly affects the maximum nonlinear phase shift that can be stabilised in a cavity: a higher modulation depth thus permits a larger stable mode-locking region of parameter space.

\subsubsection{Polarisation Effects}
\label{sec:pol}
While ideal single-mode fibre is isotropic and supports two degenerate orthogonal polarisation modes, in practice, small birefringence arises due to manufacturing imperfections and thermal / mechanical strains, breaking the degeneracy.
This results in randomly varying birefringence along the length of fibre and slightly different phase and group velocities for each polarisation axis, with the possibility of energy coupling between them.
For mode-locked fibre lasers, this uncontrolled birefringence can lead to a range of unpredictable and undesirable dynamics.

To achieve optimum stability, it is preferable to operate in a fixed, linear polarisation state: this can be achieved by constructing the cavity from polarisation-maintaining (PM) fibre.
Unfortunately, it is often necessary (and more simple / cost effective) to develop lasers using standard non-PM fibre. 
Thus, light is generated in both polarisation axes and the polarisation state evolves unpredictably with cavity propagation due to modal coupling between axes.

In this case, both polarisation eigenmodes can mode-lock simultaneously, leading to a time-varying output polarisation state.
One can consider energy from the output pulse being distributed between the two axes, with walk-off from the different axis group velocities potentially exceeding the pulse duration, and thus causing pulse break-up / partially coherent noise-burst operation~\cite{Horowitz1997}.
This walk-off is compensated under certain conditions, however, by a nonlinear trapping phenomena due to cross-phase modulation between axes, resulting in stable vector solitons~\cite{Menyuk1987a, Collings2000}.
Secondary effects can also result from random polarisation changes if polarisation-dependent loss is present in the cavity: e.g. changing the bias point (i.e. saturation properties) of an NPE-based saturable absorber or inducing spectral filtering due to the wavelength dependence of birefringence---which can vary the output pulse properties for a fixed cavity design~\cite{Kobtsev2014b}.

For the scope of this review, we consider these polarisation effects as a destabilising factor from a coherent steady-state but note that they do not directly dictate the dispersion-engineered operating regime of the cavity.
Indeed, the dynamics we discuss in Section~\ref{sec:designs} have been observed in both PM and non-PM cavities, and it is typical for non-PM mode-locked cavities to include a polarisation controller, comprising multiple waveplates, which are empirically adjusted to compensate for cavity birefringence to achieve the best possible stability.

\subsubsection{Other Nonlinear Instabilities}
\label{sec:nl_instab}
Mode-locked fibre lasers are highly attractive to the nonlinear wave dynamics research community as a testbed for exploring complex nonlinear wave phenomena and instabilities, with analogies to other physical systems.
In recent years, this has enabled observations of a range of novel phenomena including optical turbulence, soliton explosions, extreme events and rogue waves~\cite{Kartner1999,Lecaplain2012,Turitsyna2013,Runge2015,Churkin2015,Woodward_2016_pre}.
In the context of high-performance mode-locked oscillators, however, such phenomena are instabilities that need to be avoided.

One mechanism which has been suggested to fundamentally underpin destabilisation at high pulse energies is stimulated Raman scattering.
Aguergaray et al. demonstrated that the transition from stable coherent pulse generation to noise-like pulsing was accompanied by the growth of a Stokes signal, centred at the Raman gain peak~\cite{Aguergaray2013a}.
This Stokes signal was filtered out each cavity round-trip and hence was continually seeded stochastically by noise.
Nonlinear mixing between coherent pulse and incoherent Stokes light could thus lead to dephasing of the main pulse.
Further work is still required to fully understand this phenomena, and other novel nonlinear instabilities, to clarify their associated performance limitations.

\subsection{Numerical Simulations}
Numerical modelling is a valuable technique to accurately predict the output properties of mode-locked laser designs, in addition to providing insight into the intracavity dynamics.
Briefly, there are two commonly used approaches to this problem.
In the first method: a distributed average model is employed with a single equation (e.g. the Haus Master Equation / Ginzburg-Landau Equation and variations thereof~\cite{Haus2000}) to describe the whole cavity, often expressed in normalised units with input parameters representing cavity-averaged properties. 
This assumes small changes to the pulse per round trip, which may not always be accurate, although does have the benefit of permitting analytical solutions subject to various approximations.

Alternatively, models can be constructed for each cavity component with the pulse field numerically propagated through each in turn until the simulation converges to a steady-state.
This piecewise approach can be more computationally demanding, although has the benefit of clearly resolving the pulse dynamics through the cavity.
Consequently, we employ this technique (described in detail in the Appendix) extensively in Section~\ref{sec:designs}.

\section{Mode-Locked Fibre Laser Designs}
\label{sec:designs}
\begin{figure}[tb]
	\centering
	\includegraphics[width=\columnwidth]{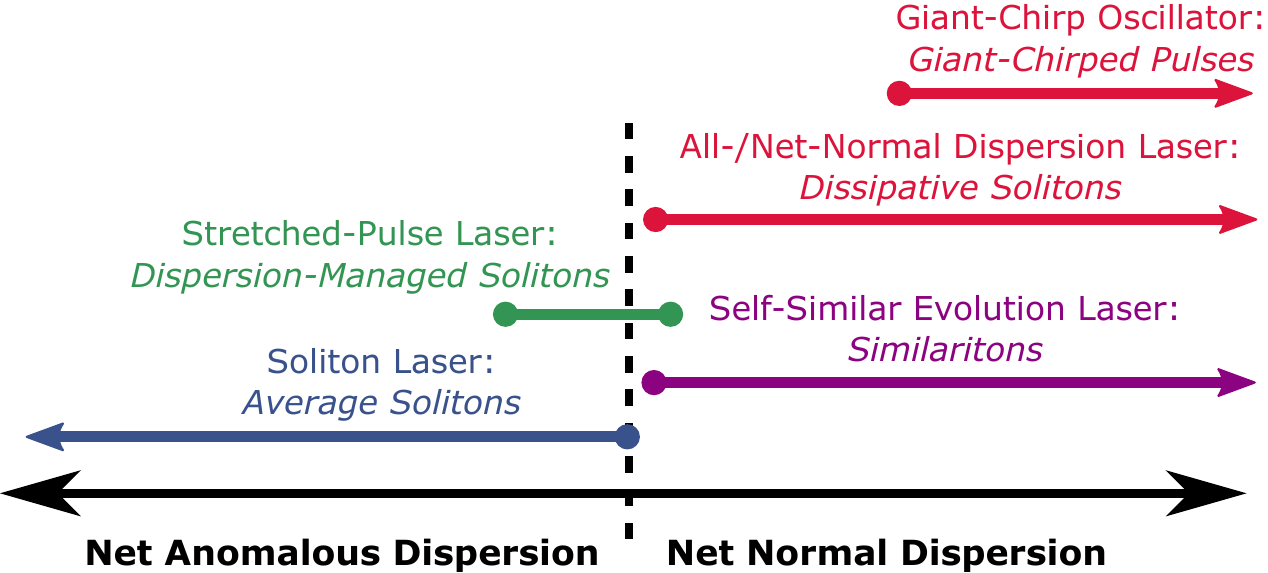}
	\caption{Summary of mode-locked fibre laser operating regimes sorted by net dispersion, including the widely used nomenclature for naming the generated pulses.}
	\label{fig:disp_maps}
\end{figure}

In this section we review the main distinct pulse shaping regimes (summarised in Fig.~\ref{fig:disp_maps}), considering their  intracavity pulse dynamics, typical reported performance and limitations.
Numerical simulations are also presented to highlight the characteristic features of each operating regime.

\subsection{Historical Perspective}
Shortly after the demonstration of rare-earth-doping of low-loss single-mode fibres and CW fibre lasers, researchers combined neodymium-doped fibre cavities (lasing at 1~\mumN) with active modulators to achieve mode-locked pulses with hundreds of picoseconds to nanosecond durations~\cite{Alcock1986,Geister1988,Duling1988}.
These pulse durations were a similar order of magnitude to the modulator switching time and dispersion played a minimal role on pulse shaping.
Improved performance was later observed with erbium-doped fibres~\cite{Hanna1989}, operating at 1.55~\mumN, due to lower fibre loss and the presence of anomalous dispersion; by including a long length of negative GVD fibre in an Er:fibre ring cavity, mode-locked pulses with sub-5~ps durations were generated, attributed to solitonic pulse shaping~\cite{Kafka1989}.
The emergence of NPE~\cite{Stolen1982b,Hofer1991} and NOLM/NALM~\cite{Doran1988, Fermann1990, Duling1991, Duling1994} techniques in the early 90s, functioning as ultrafast broadband saturable absorbers, greatly simplified cavity designs by removing active modulators and the generation of ultrashort pulses became commonplace through soliton-based mode-locking.

\subsection{Soliton Lasers}
\label{sec:desgin_sol}

Early passively mode-locked fibre lasers at 1.55~\mum were often comprised of all anomalously dispersive fibres.
In these cases, strong soliton pulse shaping effects were observed from interplay between GVD and SPM, producing ultrashort pulses (durations shorter than a few ps)~\cite{Duling1991,Duling1994,Matsas1992,Noske1992a,Tamura1992}.
As introduced in Section~\ref{sec:solitons}, periodic gain/loss and the possibility of the cavity containing fibres with different properties result in the steady-state pulse being an `average soliton', with characteristics determined by the net cavity-averaged dispersion and nonlinearity values~\cite{Kelly1991b, Hasegawa1991}.
Even for cavities including short normally dispersive sections, average soliton dynamics are maintained providing the net dispersion is sufficiently anomalous.
Average solitons are static solutions and thus, during a cavity round-trip the pulse properties only vary in amplitude due to gain/loss---the temporal and spectral widths are largely unchanged.

\begin{figure}[tbp]
	\centering
	\sffamily
	\includegraphics[width=\columnwidth]{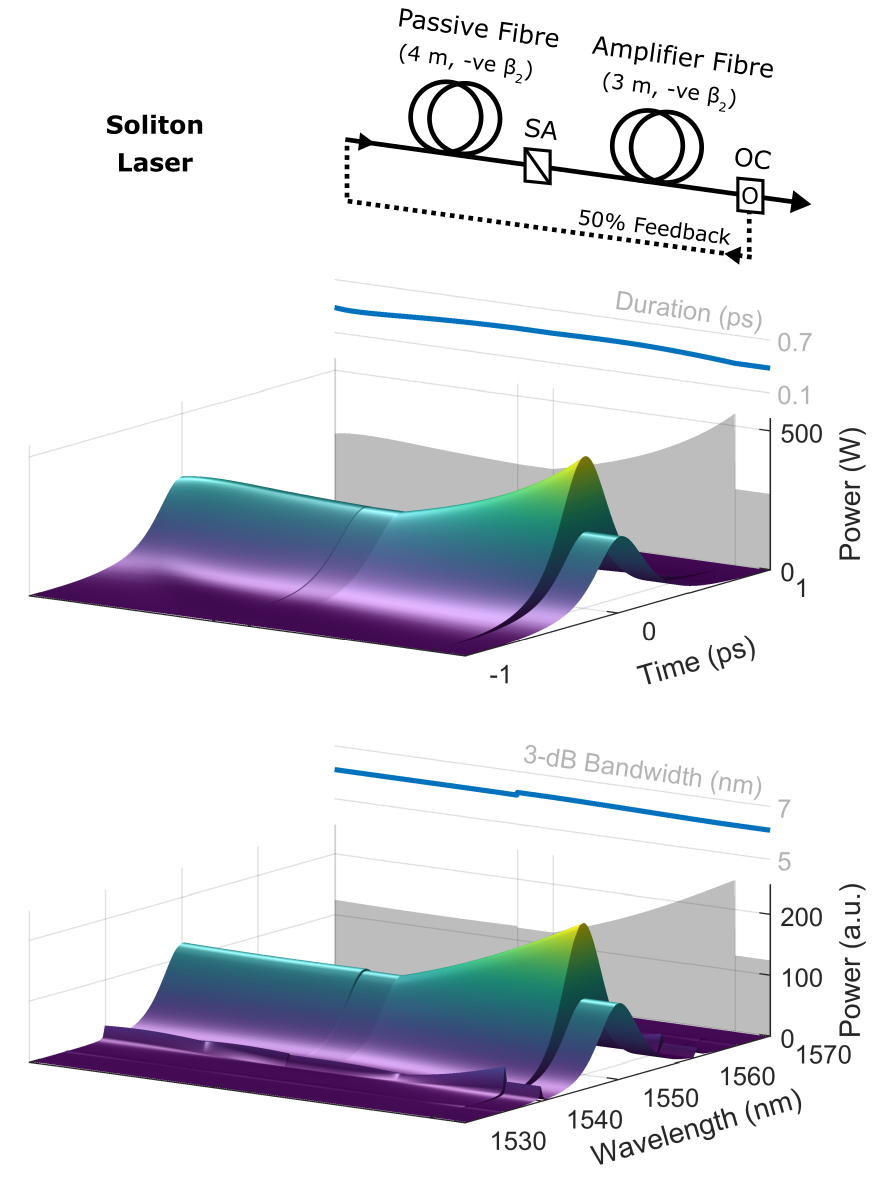}
	\begin{overpic}[width=\columnwidth]{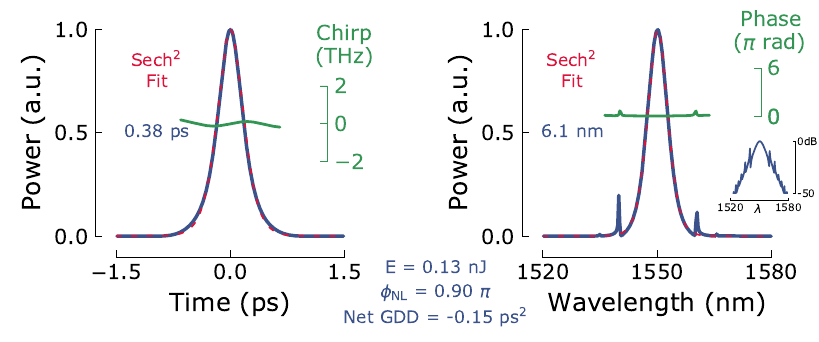}
		\put(0, 170){ {\small (a)}}
		\put(0, 40){ {\small (b)}}
		\put(47, 40){ {\small (c)}}	
	\end{overpic}
	\rmfamily
	\caption{Soliton laser: (a) cavity schematic \& steady-state round-trip pulse evolution (variation in pulse duration and bandwidth during cavity propagation are also shown, where the horizontal axis corresponds to position in the above cavity schematic); (b) output pulse profile and (c) spectrum (inset: log scale).}
	\label{fig:sol_evol}
\end{figure}

To visualise the intracavity dynamics, we simulate a typical soliton laser.
A ring cavity with 3~m gain fibre, 4~m passive fibre, a fast saturable absorber (SA) and 50\% output coupler (OC) is assumed, with fibre properties: $\beta_2=-22$~\pskm and $\gamma=1.3$~\Wkm at 1550~nm (similar to standard single-mode telecommunications fibre).
The net cavity group delay dispersion (GDD) is -0.15 ps$^2$.
Fig.~\ref{fig:sol_evol}a shows the steady-state pulse's transit through a complete cavity round trip, including the variation in pulse duration (full-width at half-maximum, FWHM) and 3-dB bandwidth.
The plotted evolutions correspond to propagation in each cavity component shown in the schematic above.
Note that the cavity schematic and evolution plots are aligned horizontally on the page such that the properties of the pulse during the round-trip (i.e. the pulse visualisation, FWHM duration and 3-dB bandwidth value) are shown directly beneath the corresponding position in the cavity schematic.
Point-action components like the SA are presented with a finite length on the propagation axis (the properties shown in this region correspond to the pulse after the component) to more clearly show their effect (i.e. the propagation axis is not to scale).
Figs.~\ref{fig:sol_evol}b--c present the laser output properties (taken after the 50\% OC).

The simulation generates 380~fs output pulses that are effectively transform-limited, with the characteristic hyperbolic secant temporal and spectral shape expected for a soliton.
Additionally, while the pulse intensity varies during the round trip, Fig.~\ref{fig:sol_evol}a confirms that changes in temporal and spectral widths/shapes are negligible (shown by the lines of nearly constant value throughout propagation in the duration and 3-dB bandwidth axes above each 3D evolution plot).

Fig.~\ref{fig:sol_evol}c highlights an additional common observation in soliton lasers: spectral (``Kelly'') sidebands~\cite{Kelly1992}.
Periodic perturbations to the pulse in the cavity result in reshaping to maintain the soliton shape; in doing this, the soliton sheds energy as linear dispersive radiation over a broad spectrum.
At certain frequencies, the dispersive radiation propagating in the cavity is phase-matched to the soliton, resulting in resonant enhancement of the dispersive wave to form sidebands in the spectrum.
Temporally, this energy results in a broad pedestal under the pulse.
The effect becomes stronger as the ratio of cavity length to soliton period $L/Z_0$ increases and can result in high-intensity sidebands comprising a large portion of the laser output power.
In this case, the actual peak power of the soliton is low and applications of the laser will be limited.
Practically, this limits the shortest possible pulse duration (as $Z_0\propto\tau^2$).
Even if $L<<Z_0$ is satisfied, the pulse energy is still limited by the soliton area theorem based on the cavity-averaged dispersion and nonlinearity values.

The avoidance of the nonlinear instability can also be expressed as the need to minimise the pulse's nonlinear phase accumulation per round-trip (also known as the B integral):
\begin{equation}
\phi_\mathrm{NL} = \frac{2\pi}{\lambda}\int_{0}^{L}\frac{n_2 P(z)}{A_\mathrm{eff}} dz
\end{equation}
where $P(z)$ is the peak power.
For soliton lasers, $\phi_\mathrm{NL}$ should be much less than $2\pi$, since this corresponds to the resonant instability case~\cite{Mollenauer1986,Kelly1992}.
We note that our simulated 0.13~nJ energy soliton laser still shows modest sidebands for a value of $\phi_\mathrm{NL}=0.9\pi$.

Soliton mode-locking has been widely demonstrated with ytterbium~\cite{Okhotnikov2003,Lim2003,Avdokhin2003}, erbium~\cite{Duling1991,Duling1994,Matsas1992,Noske1992a,Tamura1992} and thulium~\cite{Nelson1995, Sharp1996} gain fibre in the near-infrared, using dispersion compensation to introduce large anomalous dispersion for regions below the zero dispersion wavelength of conventional fibre.
Performance has been limited to 100s fs pulse durations and 10s--100s pJ energies, however, due to fundamental soliton restrictions outlined above.

\subsection{Stretched-Pulse Lasers}
\label{sec:desgin_dm_sol}
The limitations of soliton lasers can be mitigated by introducing a dispersion map: i.e. sections of alternating signs of GVD, yielding a near-zero net cavity dispersion.
Some of the earliest mode-locked lasers achieved this by employing normally dispersive fibre alongside bulk components such as prisms or diffraction gratings to introduce anomalous dispersion~\cite{Ober1993}. 
The concept was formalised and extended by Tamura et al.~\cite{Tamura1993} to become the \textit{stretched-pulse laser} design, using alternating segments of normally and anomalously dispersive fibre to cause the circulating pulse to periodically broaden and compress.
The `breathing' pulse solution can change duration by over an order of magnitude, corresponding to a reduction in the averaged value of pulse peak power around one cavity round trip compared to a conventional soliton laser: thus permitting higher pulse energies before the onset of multi-pulsing / instabilities.
Pulses in this operating regime are often referred to as \emph{dispersion-managed solitons} (analogous to solitons in dispersion-managed optical fibre transmission lines~\cite{Lin1980}).

\begin{figure}[tbp]
	\centering
	\sffamily
	\includegraphics[width=\columnwidth]{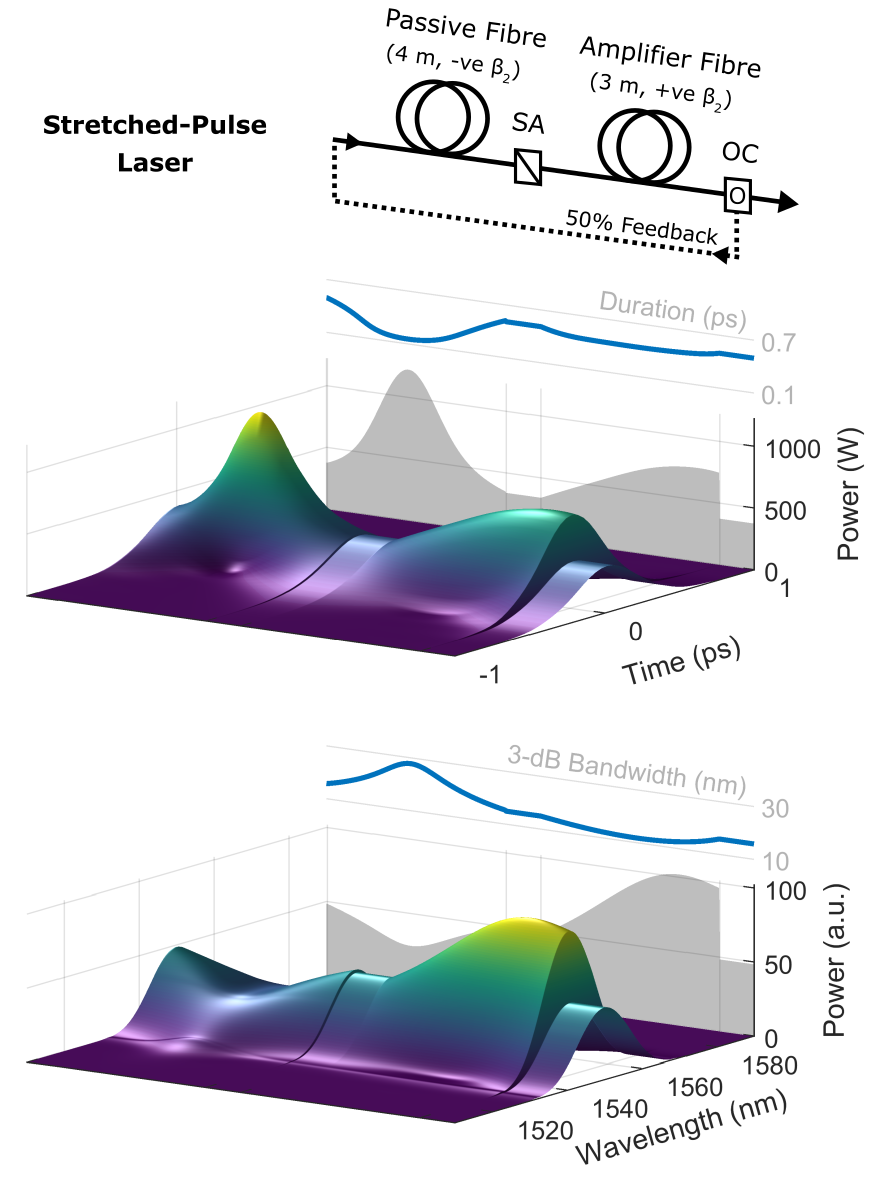}
	\begin{overpic}[width=\columnwidth]{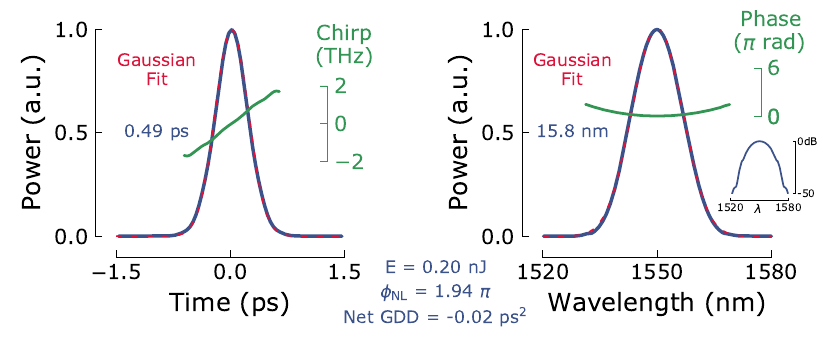}
		\put(0, 170){ {\small (a)}}
		\put(0, 40){ {\small (b)}}
		\put(47, 40){ {\small (c)}}	
	\end{overpic}
	\rmfamily
	\caption{Stretched-pulse laser: (a) cavity schematic \& steady-state round-trip pulse evolution; (b) output pulse profile and (c) spectrum (inset: log scale).}
	\label{fig:dm_sol_evol}
\end{figure}

A typical steady-state dispersion-managed soliton round-trip is shown in Fig.~\ref{fig:dm_sol_evol}.
This simulation is based on the same cavity as in Section~\ref{sec:desgin_sol}, but with the sign of gain fibre's GVD inverted.
The net cavity dispersion is -0.02~ps$^2$.
Compared to Fig.~\ref{fig:sol_evol}, the stretched-pulse laser exhibits clear breathing in a round trip: here, the pulse duration varies from 155~fs to 520~fs (with bandwidths in the range 12--28~nm).
The minimum pulse width is reached in the middle of each fibre (at these points, the pulse is transform-limited), whereas at the ends of each section the pulse is broadened and strongly chirped.
In this simulation, the pulse is extracted after the normally dispersive fibre section and is up-chirped with a time-bandwidth product of 0.97. 
Due to the linearity of the chirp, extra-cavity compression back to the transform-limit (220~fs) is possible.
It should also be noted that unlike sech$^2$-shaped solitons described earlier, the temporal and spectral pulse shape for stretched-pulse lasers are Gaussian~\cite{Nelson1997}.
Additionally, the simulation permitted higher pulse energies before the onset of multi-pulsing: Fig.~\ref{fig:dm_sol_evol} shows a stable steady-state pulse with 0.2~nJ energy.

The steady-state pulse is stabilised by a balance between nonlinear and dispersive phase, similar to soliton lasers, although dispersion-managed soliton operation is not limited to cavities with net anomalous dispersion.
Indeed, stable pulse generation is supported for net-zero and even small net-normal dispersion values.
While the observation of soliton dynamics for cavity-averaged positive dispersion values may seem counter-intuitive, this can be explained by the change in spectral bandwidth during a round-trip: in the anomalous dispersion sections, the bandwidth is larger, thus anomalous GVD segments have a greater influence on phase than normal GVD sections, since spectral phase accumulation is proportional to the product of GVD and the square of the frequency offset from the pulse centre frequency.
It should also be noted that a highly unbalanced dispersion map with strong net anomalous dispersion results in the operating regime becoming average solitonic (Section~\ref{sec:desgin_sol}) with the associated limitations described earlier~\cite{Nelson1997}.

Dispersion management enables higher pulse energies to be supported and the varying cavity dispersion also disturbs phase matching of dispersive waves (emitted during soliton shaping sections), which minimises the resonant sideband instability~\cite{Tamura1994}.
Therefore, spectral sidebands are characteristically absent in the spectra of stretched pulse lasers, permitting greater nonlinear phase accumulation without destabilisation.
Since the coupling between GVD and frequency deviations leads to pulse timing jitter in a mode-locked laser, the reduced cavity dispersion of stretched-pulse lasers also yields lower timing jitter\cite{Namiki1997}.

In terms of stretched-pulse laser design and dispersion engineering, the evolving nature of the pulse in the cavity results in the position of the output coupler and ordering of components affecting the output pulse properties. 
For example, extracting the pulse after the anomalously dispersive fibre would result in a down-chirped pulse.
Pulse with the broadest possible spectra (with shortest compressed pulse durations) and highest pulse energy in this regime are generated for net-zero / small net-normal cavity GDD, with strong contributions from SPM to spectral broadening through nonlinear phase shifts~\cite{Nelson1997}.

Stretched-pulse laser designs significantly improve upon the performance of strongly net-anomalous soliton lasers, routinely enabling the generation of up to nanojoule-level pulses with 100s~fs to few~ps durations, compressible to $<$100~fs~\cite{Ober1993, Tamura1993, Nelson1996, Zhang2013l,Nomura2014a}.
Near gain-bandwidth-limited spectral widths have also been achieved, enabling extracavity compression to sub-50~fs durations throughout the near-IR~\cite{Ober1993,Zhang2013l,Nomura2014a}.
For detailed further analysis of this regime, Refs.~\cite{Nelson1997,Turitsyn2012} are recommend.
Despite improved performance, the power scalability is still limited by the onset of nonlinear multi-pulsing instabilities, prohibiting pulse energies exceeding a few nJ.

\subsection{All-Normal \& Net-Normal Dispersion Lasers}
In the 1990s, the role of dispersion in mode-locked cavities had been widely studied and it was well-known that highly chirped pulses are solutions of the Haus Master Equation for large normal dispersion~\cite{Haus1992}.
However, as the field of passively mode-locked fibre lasers was emerging, it was suggested that appreciable normal dispersion from just a few metres of fibre would result in continually broadening temporal envelopes due to a lack of stabilisation mechanism (without solitons), preventing self-starting pulse generation~\cite{Hofer1992}.
A small number of demonstrations did achieve stable mode-locking without dispersion compensation (for Yb and Nd lasers operating at 1~\mumN), although they used carefully engineered short lengths of doped fibre to minimise the net dispersion~\cite{Hofer1992, Orsila2004}.

In 2006, Chong et al. reported an advance in the concept of dispersion engineering of fibre lasers by introducing the all-normal dispersion (ANDi) architecture, with spectral filtering highlighted as an important factor in the cavity dynamics to stabilise the pulse, even in highly normally dispersive environments\cite{Chong2006}.
There had been a small number of prior demonstrations of stable mode-locking in strongly net-normal GVD mode-locked fibre lasers~\cite{Zhao2006b, Zhao2006c, DeMatos2004} (e.g. with sufficient filtering provided by the limited gain profile, known as `gain-guiding'~\cite{Zhao2006b}), although this work was important in instigating a shift towards higher energy mode-locked fibre lasers by eliminating anomalous sections, and uncovering the underlying nonlinear dynamics~\cite{Chong2008a}.
Here, solitonic shaping is absent and the interaction of linear and nonlinear effects leads to a positive monotonic frequency sweep (up-chirp) across the pulse.
Therefore, the generated pulses are compressible outside the cavity.

\begin{figure}[tbp]
	\centering
	\sffamily
	\includegraphics[width=\columnwidth]{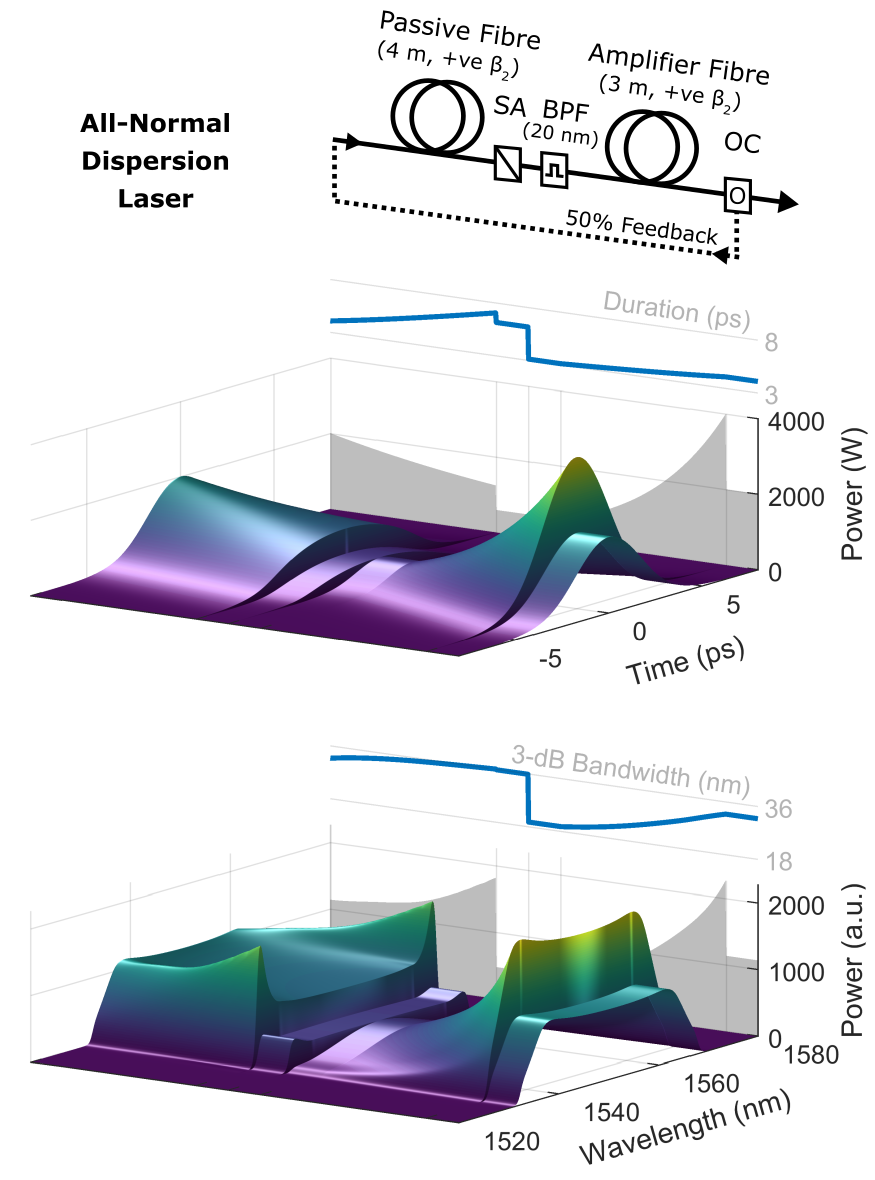}
	\begin{overpic}[width=\columnwidth]{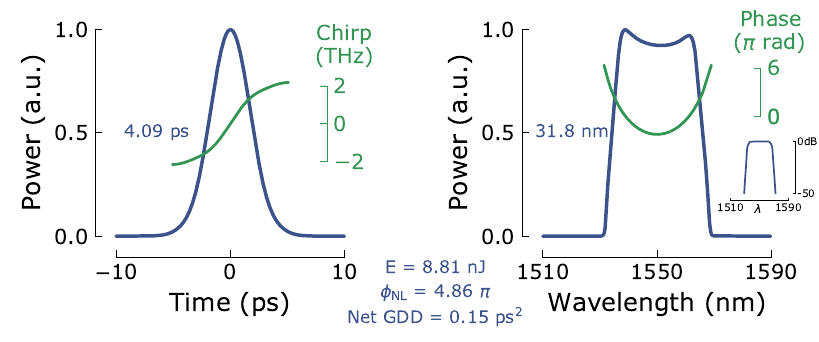}
		\put(0, 170){ {\small (a)}}
		\put(0, 40){ {\small (b)}}
		\put(47, 40){ {\small (c)}}	
	\end{overpic}
	\rmfamily
	\caption{All-normal dispersion laser: (a) cavity schematic \& steady-state round-trip pulse evolution; (b) output pulse profile and (c) spectrum (inset: log scale).}
	\label{fig:andi_evol}
\end{figure}

In ANDi lasers, both dissipative (spectral filtering, gain and loss) and conservative (dispersive and nonlinear) effects play a major role in the pulse shaping process, unlike in anomalous dispersion soliton lasers which rely principally on a balance between GVD and SPM (gain and loss are of course present, but less significant to pulse shaping). 
These pulses have therefore become known as \textit{dissipative solitons}~\cite{Akhmediev2005, Grelu2012}.
Specifically, the pulse broadens as it propagates along normally dispersive fibre and SPM results in spectral broadening. 
After each round-trip the saturable absorber decreases the pulse width and a spectral filter limits the pulse bandwidth, thus balancing the broadening in each round trip.
Due to the chirped nature of the pulse, temporal (spectral) filtering also has a spectral (temporal) width reduction effect.
In terms of the cavity amplitude and phase balance, GVD acts to linearise the nonlinear phase accumulation from SPM, and linear phase accumulation is balanced by spectral filtering and saturable absorption, which also restricts the pulse amplitude~\cite{Renninger2012}.

These dynamics are visualised for a typical ANDi laser simulation in Fig.~\ref{fig:andi_evol}.
Compared to the previous simulations, all cavity fibre is normally dispersive here (net GDD $=0.15$ ps$^2$) and a bandpass filter (BPF, Gaussian shape with 20~nm FWHM) is added before the amplifier.
The temporal and spectral shape vary throughout the cavity, but compared to the dispersion-managed soliton evolution, the breathing ratios are smaller (breathing ratio $\sim2$) and widths are not slowly continuously varying: the dissipative soliton experiences instantaneous reductions in pulse width and bandwidth due to filtering. 
Such dissipative processes and SPM lead to characteristic steep-sided optical spectra, and `cat ears' or `M-shaped' spectral peaks~\cite{Chong2008a}.
Our simulated ANDi laser generates strongly chirped $\sim$8.8~nJ pulses with 4.1~ps duration and 31.8~nm bandwidth (compressible to $<$120~fs). 
By eliminating pulse shaping in anomalous dispersion sections, the ANDi laser avoids the sideband / area theorem limitation of classical solitons and enables significantly higher pulse energies.
It has been empirically determined that the maximum nonlinear phase shift supported by a dissipative soliton is $\sim$10$\pi$~\cite{Renninger2015}.

The ANDi architecture simplifies mode-locked laser designs at 1~\mum based on highly efficient ytterbium amplifiers by removing the need for dispersion compensation.
Consequently, many Yb:fibre mode-locked lasers followed, with impressive pulse energies exceeding 10~nJ and compressed sub-200~fs durations~\cite{Kieu2009, Chong2007, An2007, Ruehl2008} (notably 31~nJ in Ref.~\cite{Kieu2009}), corresponding to peak powers above 100~kW.
Further scaling to the near-microjoule level has also been demonstrated using rod-type fibres~\cite{Ortac2009}, although we restrict our focus here to typical single-mode fibre lasers. 

To extend this laser design to longer wavelengths (beyond the silica zero-dispersion wavelength), it is important to question the influence of anomalous dispersion in the cavity---i.e. what happens for a large net-normal (but not \emph{all}-normal) cavity dispersion?
The addition of short anomalous GVD sections to ANDi cavities has been found to have only minor influence: dissipative soliton dynamics and typical characteristics are maintained~\cite{Zhao2006c}.
However, a qualitatively different pulse evolution is introduced for a net-normal dispersion map including strong anomalous dispersion.
While the potential for classical solitonic shaping in the anomalous segment can obstruct the formation of stable high-energy dissipative solitons~\cite{Tang2015b}, an amplitude/phase balance is possible in which the pulse breathes continuously due to dispersive broadening / compressing~\cite{Ilday2003b, Renninger2012}.
Analogous to dispersion-managed soliton shaping, such pulses have been labelled `stretched dissipative solitons' within a recent excellent review of normal-dispersion pulse shaping\cite{Renninger2012}, although the dynamic remains relatively unexplored, with further work needed to understand and exploit it.

With a focus on minimising anomalous dispersion, dissipative solitons at the 10~nJ level have been generated in net-normal dispersion cavities at erbium~\cite{Ruehl2008, Jeong2014b,Tang2015c} (notably 38~nJ in Ref.~\cite{Tang2015c}) and thulium wavelengths~\cite{Tang2015b, Gaponov2015, Huang2015b}.
While the dissipative soliton pulse evolution enables significantly enhanced energies, scalability is still limited by the onset of multi-pulsing by wave breaking, excess nonlinear phase accumulation and saturable absorber overdriving.

\subsubsection{Long-Cavity Giant-Chirp Oscillators}
The pulse energy of a mode-locked laser is inversely proportional to the repetition rate and thus, directly related to cavity length.
A simple approach to energy scaling is, therefore, cavity elongation. 
This concept was previously tested in many of the early mode-locked soliton fibre lasers with 100s~m cavity lengths, although it was found that elongation was accompanied by a tendency for incoherent, noise-burst operation (linked to the soliton sideband instability discussed earlier)~\cite{Putnam1998}.
ANDi lasers do not suffer from such length limitations, however, and as understanding of dissipative soliton dynamics developed, the concept of a `giant-chirp oscillator' emerged: by using normally dispersive fibre to lengthen the cavity, coherent pulse generation could be preserved, with significant pulse broadening and a large linear chirp from the additional dispersive phase~\cite{Renninger2008a}.

\begin{figure}[tbp]
	\centering
	\sffamily
	\includegraphics[width=\columnwidth]{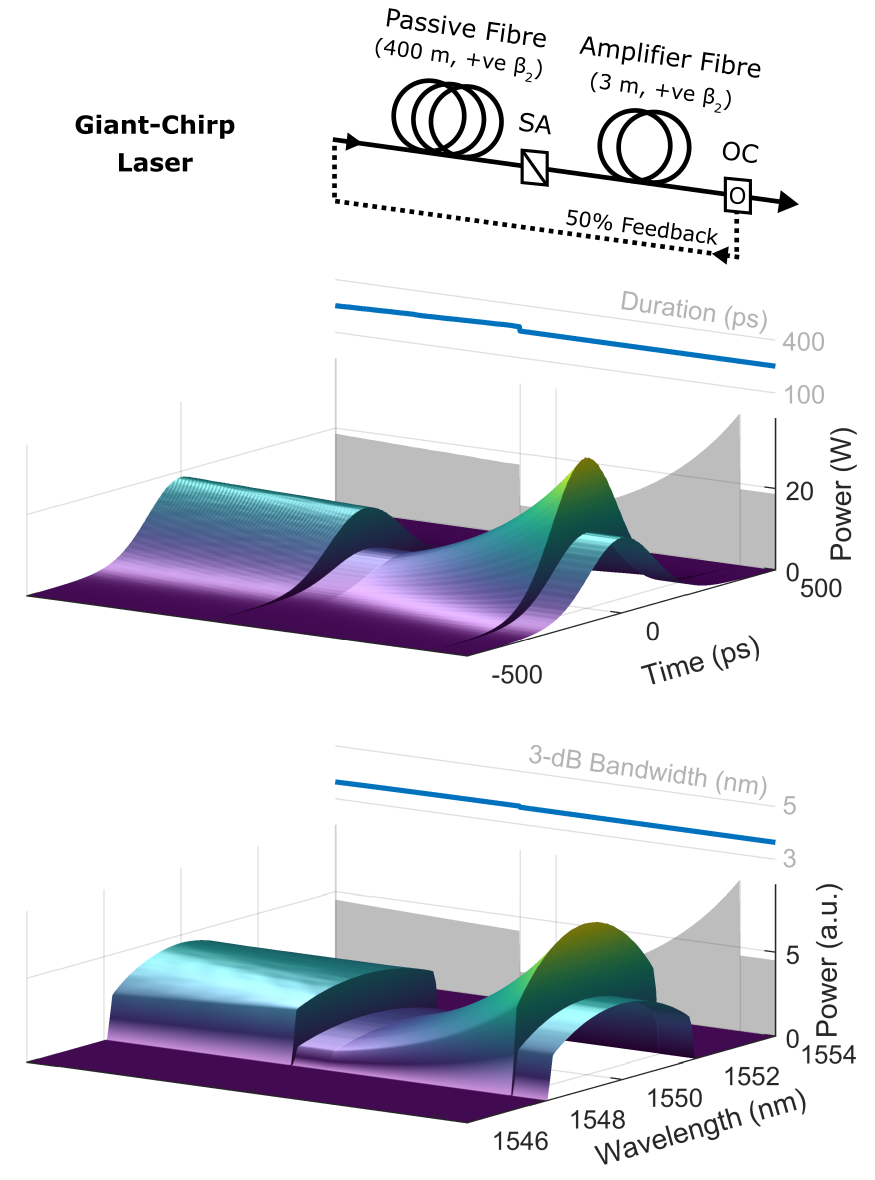}
	\begin{overpic}[width=\columnwidth]{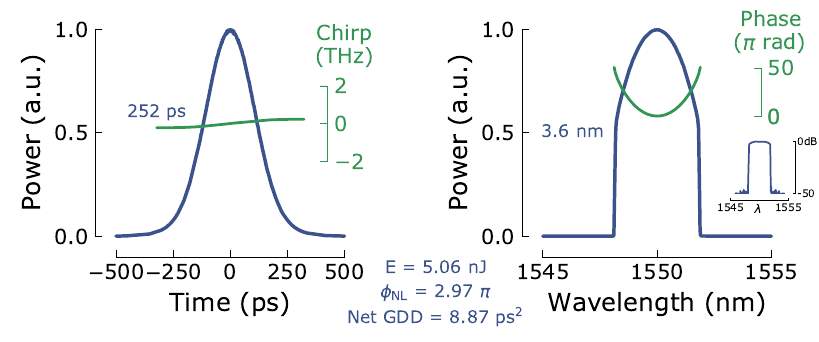}
		\put(0, 170){ {\small (a)}}
		\put(0, 40){ {\small (b)}}
		\put(47, 40){ {\small (c)}}	
	\end{overpic}
	\rmfamily
	\caption{Giant-chirp oscillator: (a) cavity schematic \& steady-state round-trip pulse evolution; (b) output pulse profile and (c) spectrum (inset: log scale).}
	\label{fig:gco_evol}
\end{figure}

We simulate a typical giant-chirp oscillator by extending the passive fibre in our ANDi cavity to 400~m (Fig.~\ref{fig:gco_evol}).
Stable coherent pulses with 252~ps width are generated---significantly longer than the few ps pulses from a typical mode-locked laser, due to the high net cavity GDD of +8.9~ps$^2$---with a predominantly quadratic phase and thus, linear chirp.
The broad pulses possess low peak power, however, resulting in minimal spectral broadening through SPM and thus, a relatively narrow 3.6~nm spectrum.
In terms of pulse dynamics and the steady-state energy balance, giant-chirp oscillators can be considered a subset of linearly chirped dissipative soliton ANDi lasers.
However, for the giant-chirp pulse evolution, the gain-guiding filtering effect was sufficient to stabilise the pulse; thus, a bandpass filter was not explicitly required here.
Changes in temporal and spectral shape during each round-trip are also negligible, with a breathing ratio of $\sim$1.

Despite the giant nature of the chirp, it has been shown that compression using standard grating dispersion compensation (i.e. linear compensation) can result in compression to pulse durations within a factor of two of the transform limit~\cite{Renninger2008a}.
Extension up to 100~m cavity lengths (few MHz repetition rates) has enabled 10s--100s~ps pulse generation with up to $\sim$15~nJ energies, compressible to 100s~fs~\cite{Erkintalo2012,Renninger2008a}.
The real benefit of this architecture, however, is in simplifying amplification systems---broad high-energy chirped pulses at a low repetition rate from the oscillator can directly replace the typical chain of seed oscillator, pulse picker, stretcher and pre-amplifiers, enabling direct amplification to microjoule energies without distortion~\cite{Renninger2008a}.
However, energy scalability has been shown to be limited by the onset of intracavity Raman scattering at high peak powers~\cite{Aguergaray2013a}.

This concept has also been extended to kilometre length scales to produce giant-chirped nanosecond pulses with 100s~kHz repetition rates~\cite{Kelleher2009,Tian2009,Zhang2009d,Woodward_ptl_2014}.
Such broad pulses are typically accompanied by narrow ($<$1~nm) spectra due to the low intracavity peak powers yielding reduced SPM broadening---thus raising barriers to compression by the need for carefully engineered dispersion compensation.
With bespoke chirped FBGs, however, compression by two orders of magnitude from $\sim$1~ns to few ps durations has been demonstrated~\cite{Woodward_ol_2015_gco}.
Recent work has also considered the concept of nonlinearity management: carefully positioning long fibre sections to exploit, but not overdrive, nonlinear effects: achieving sub-MHz repetition rates with bandwidths supporting grating-based compression to $<$300~fs~\cite{Bowen2016b}.

\subsection{Similariton Lasers}
In 1993, Anderson et al.\ demonstrated that high-energy pulse propagation in normally dispersive fibre could avoid wave breaking for pulses with a quadratic temporal intensity profile (i.e. parabolic shape) and quadratic temporal phase (i.e. linear chirp), which were asymptotic solutions to the NLSE~\cite{Anderson1993}.
With increasing propagation distance, the spectral and temporal width of a parabolic pulse increase, but the shape is maintained---i.e. the pulse evolves self-similarly, leading to the name \textit{similariton}.
(For further details on self-similarity in ultrafast photonics, Ref.~\cite{Dudley2007} is recommended.)
It was later shown numerically~\cite{Tamura1996} and experimentally~\cite{Fermann2000} that even arbitrarily shaped pulses injected into normal-dispersion fibres would asymptotically evolve towards this parabolic, chirped solution, providing they were experiencing gain (i.e. in a fibre amplifier).
Chirped parabolic pulses were later analytically shown to be a nonlinear attractor in normally dispersive gain fibre~\cite{Fermann2000}.
To rigorously distinguish between the self-similar pulse evolution in passive fibre, these pulses are often referred to as \textit{amplifier similaritons}.

Similaritons present many attractive features, including the avoidance of wave breaking and no limits to the nonlinear phase accumulation~\cite{Chong2015}.
The pulse quality of parabolic intensity distributions is excellent, with minimal pulse wings, and the highly linear chirp permits compression to the transform-limit without distortion.
Consequently, an active research effort has explored the generation of similaritons from mode-locked fibre lasers---effectively considering self-similar pulse propagation subject to periodic boundary conditions~\cite{Chong2015}.

While, in principle, any pulse injected into a normal-GVD gain fibre will evolve towards the parabolic pulse attractor, the length scale to achieve this depends strongly on the input condition.
Thus, to generate similaritons from practical fibre amplifier lengths, the input should be a near-transform-limited pulse~\cite{Chong2015}.
Amplifier similariton evolution can be exploited as a mode-locked laser dynamic by permitting the pulse to evolve towards the attractor solution each round-trip (with temporal and spectral broadening), then coupling out a portion of light and `resetting' the properties of the field (reducing the duration and bandwidth) that is fed back for the next iteration.
Under such conditions, with a saturable absorber to initiate mode-locking, steady-state similariton mode-locking can be achieved.

Various approaches have been demonstrated to achieve pulse resetting, including soliton re-shaping in an anomalously dispersive fibre segment~\cite{Oktem2010} and strong spectral filtering~\cite{Renninger2010a, Bale2010}.
Alternatively, the resetting requirements are relaxed by allowing the pulse to evolve slowly over a long amplifier length (e.g. 2.4~km~\cite{Aguergaray2010}) by using distributed Raman gain in place of rare-earth-doped fibre.
It should also be noted that a passive normal dispersion fibre with decreasing dispersion is equivalent to normally dispersive gain fibre in terms of pulse evolution~\cite{Hirooka2004} and `amplifier similaritons' have been observed in such passive fibre~\cite{Finot2007}.

\begin{figure}[tbp]
	\centering
	\sffamily
	\includegraphics[width=\columnwidth]{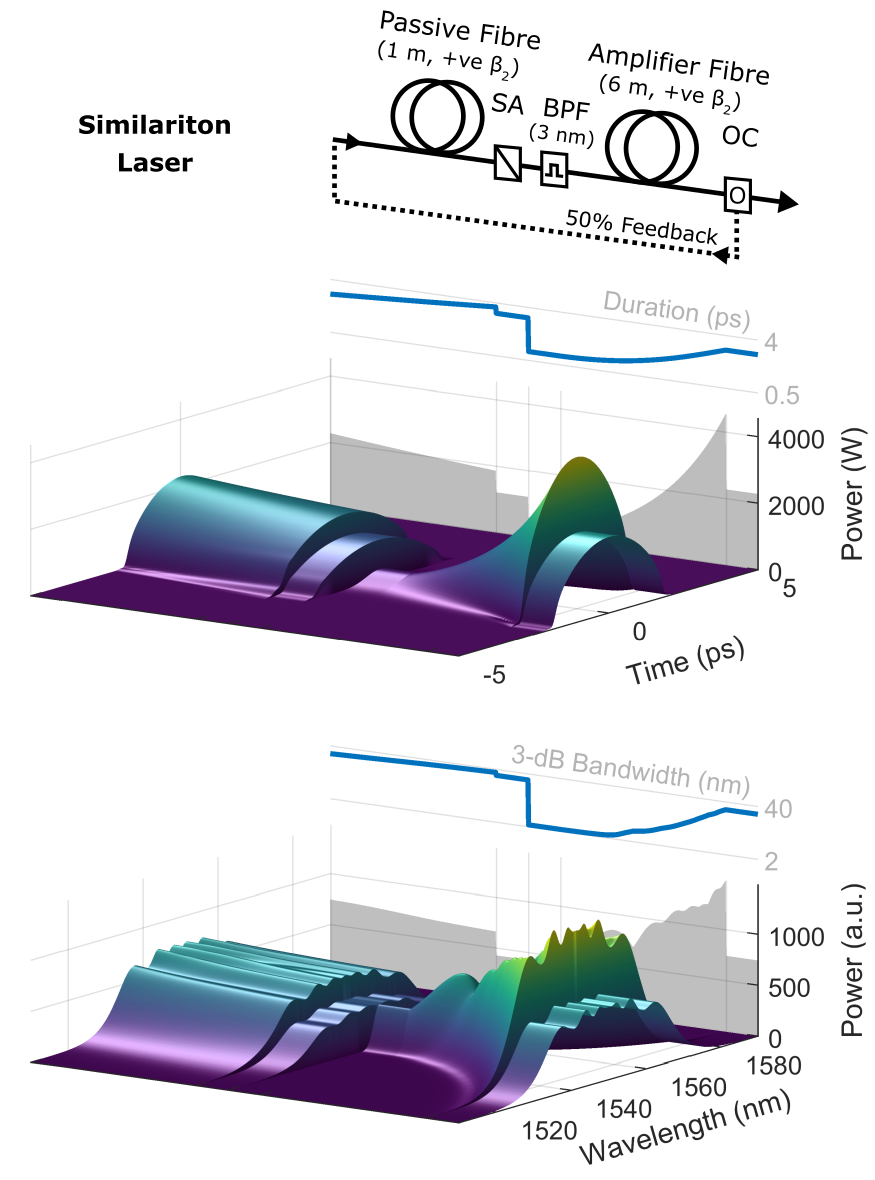}
	\begin{overpic}[width=\columnwidth]{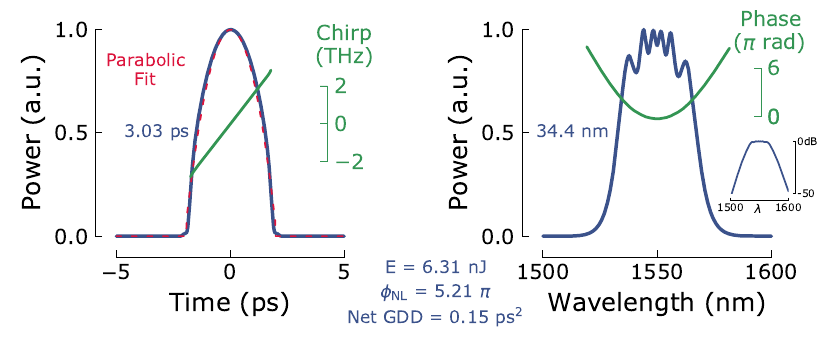}
		\put(0, 170){ {\small (a)}}
		\put(0, 40){ {\small (b)}}
		\put(47, 40){ {\small (c)}}	
	\end{overpic}
	\rmfamily
	\caption{Similariton laser: (a) cavity schematic \& steady-state round-trip pulse evolution; (b) output pulse profile and (c) spectrum (inset: log scale).}
	\label{fig:ss_evol}
\end{figure}

We simulate a laser exhibiting similariton dynamics with a cavity design akin to the ANDi laser, but modified by using a longer 6~m gain fibre and narrower 3~nm filter, shown in Fig.~\ref{fig:ss_evol}.
The 3~ps output pulses are well-fitted by a parabolic shape and exhibit a linear chirp, with 34.4~nm bandwidth (compressible to 170~fs).
The 6.31~nJ pulse energy and 5.21$\pi$ nonlinear phase shift significantly exceed the performance of anomalous-dispersion soliton lasers.
Our simulation also shows another salient feature of similariton lasers: strong breathing.
Spectral bandwidth varies by a factor of 11, with greatest bandwidth generated after the passive fibre, followed by a small reduction from the saturable absorber (due to the coupling between spectral and temporal width for chirped pulses) and a larger `reset' operation from the narrowband filter.
The corresponding temporal breathing ratio is 3.4.

The highest-performance similariton lasers to date have generally exhibited slightly lower pulse energies than dissipative soliton sources, but enabled shorter compressed pulses---e.g. 3~nJ 55~fs pulses in Ref.~\cite{Renninger2010a}, 10~nJ 42~fs pulses in Ref.~\cite{Nie2011}, both at 1~\mum and 33~nJ 156~fs pulses at 1.55~\mum~\cite{Chu2015}.
A major limitation is the limited gain bandwidth of common rare-earth dopants: similariton bandwidth scales with pulse energy as E$^{1/3}$, thus if the pulse bandwidth starts to exceed the finite gain width, the resulting unequal spectral amplification disturbs the quadratic phase, disrupting self-similar evolution~\cite{Renninger2010a}.
It should also be stressed that while similaritons and dissipative solitons can be considered as distinct mode-locking states, a wide range of intermediate regimes exist.
Numerous normal-dispersion mode-locked lasers have thus reported operation with behaviours between the characteristic properties---e.g. quasi-parabolic pulse shapes~\cite{Tang2015c,Tang2015b}.
Work to date also suggests that achieving a self-consistent similariton evolution is more challenging than for a dissipative soliton (e.g. filter bandwidth is critical~\cite{Nie2011,Bale2010,Wang2017b}).

Finally, it should be noted that it is also possible to generate chirped parabolic pulses based on an average-cavity similariton evolution, rather than relying solely on the nonlinear attractor in amplifier fibre. 
A parabolic pulse converts nonlinear phase accumulation into a linear chirp in the presence of normal dispersion~\cite{Chong2015}, which can be balanced in the net-normal dispersion cavity by employing a carefully designed dispersion map~\cite{Ilday2004}.
Reports of such \emph{passive similariton lasers} are rarer, however, and performance to date has not exceeded 10~nJ pulse energies~\cite{Ilday2004,Ruehl2005}.
Further work will hopefully improve understanding of self-similarity in laser cavity dynamics, with opportunities for higher performance pulse sources.

\section{Discussion \& Outlook}
\label{sec:outlook}

\subsection{Dispersion Engineering for High-Energy Pulse Generation}
To generate high-energy ultrashort pulses, which mode-locked operating regime is optimum?
Unfortunately, there is no clear answer.
Each cavity dynamic described in Section~\ref{sec:designs} offers various benefits, but also disadvantages for practical mode-locked laser systems; ultimately, the design should be matched to the target application.
We can, however, draw some general conclusions from recent progress in this area.

Firstly, for high-energy pulse generation, normal dispersion-based dynamics are preferable.
Net-anomalous cavities are fundamentally limited by soliton dynamics and while the stretched-pulse concept can mitigate these to an extent (with the downside of requiring a carefully designed dispersion map), their best reported performance lags behind typical normal-dispersion lasers.

ANDi / large-net-normal dispersion lasers producing dissipative solitons offer high-performance with the simplest design, including relatively weak restrictions on the bandwidth required for filtering and the overall dispersion value.
The linear chirp is simple to compensate with standard pulse compression techniques, typically resulting in few nJ femtosecond pulses at tens of MHz repetition rates.
A general feature of dissipative soliton lasers is steep spectral edges, corresponding to a sinc function in the time domain via the Fourier transform; thus even transform-limited pulses generally exhibit low-intensity oscillating pulse wings (with $\sim$5\% of the energy~\cite{Renninger2012}).
Despite this, such lasers have been widely reported, and even recently commercialised~\cite{KM2017,SP2017}.

Giant-chirp oscillators represent an interesting route to energy scaling through duty cycle enhancement, well-suited for applications requiring high-peak-powers with low average power (e.g. biomedical imaging~\cite{Krolopp2016}). 
The compressed pulse quality is often worse than shorter ANDi lasers, however.
Aside from this limitation, the design does offer simplifications for chirped-pulse amplification systems.

Similariton mode-locked lasers are a more recent development.
The design criteria to achieve stable similariton evolution appears more stringent than dissipative soliton shaping: the stabilisation mechanism to ensure periodic self-similar evolution is particularly critical.
Various studies of normal dispersion cavities have demonstrated that similariton generation occupies only a small region of parameter space; for example, by changing the bandwidth of a laser's spectral filter out of a small few-nm range, pulse dynamics transition from a similariton to dissipative soliton state~\cite{Nie2011,Wang2017b}.
When stable similariton operation is achieved, however, there are various benefits.
The generated parabolic pulses with highly linear chirp can be perfectly compressed (and the smoother spectral shape avoids a time-domain sinc function), giving excellent quality pulses.
Parabolic pulses are also ideally suited to subsequent amplification, avoiding wave-breaking phenomena.

While novel pulse dynamics can overcome certain nonlinear instabilities (e.g. wave breaking), other limitations may be more fundamental.
Stimulated Raman scattering, for example, has been shown to lead high-energy normal-dispersion lasers to exhibit noisy operation and soliton explosions~\cite{Aguergaray2013a, Runge2015}.
Other reports, however, have demonstrated stable pulsation in the presence of Raman scattering through ensuring cavity resonance for both signal and the Stokes wave (thus avoiding the noise-seeded Stokes pulse evolution each round trip)~\cite{Bednyakova2013a}.
This illustrates that a complete understanding of these various nonlinear phenomena and the limits they present to high-performance mode-locking still requires further research, although may be mitigated through careful cavity design.

Finally, it is important to note that the classification of laser regimes in Section~\ref{sec:designs} is by no means definitive.
While it is instructive to categorise dynamical operating regions to identify trends, design criteria and to understand their typical performance and limitations, these are limiting cases; in practice, a continuum of pulse behaviours exists due to the vast parameter space offered by the design flexibility of fibre lasers.
Intermediate states to (and regions which completely defy) the above classification have been reported, with novel properties (e.g. triangle-shaped pulses~\cite{Boscolo2012b}, spiny solitons~\cite{Chang2015a} etc.), highlighting that a large part of parameter space still remains unexplored and poorly understood.
It is expected that further investigation will reveal new pulse dynamics and characteristics, which could be practically exploited for high-energy ultrafast lasers.

\subsection{Automated Engineering of `Smart' Mode-Locked Lasers}
There has been significant progress recently in the development of automated techniques to control cavity parameters, potentially simplifying the design process and offering widely tunable performance from a single laser design.
Cavity dispersion, filtering and the saturable absorber response are critical to establishing the amplitude/phase balance for a particular steady-state, and recent reports have demonstrated approaches to dynamically control these variables in-situ.

\begin{figure}[tb]
	\centering
	\includegraphics{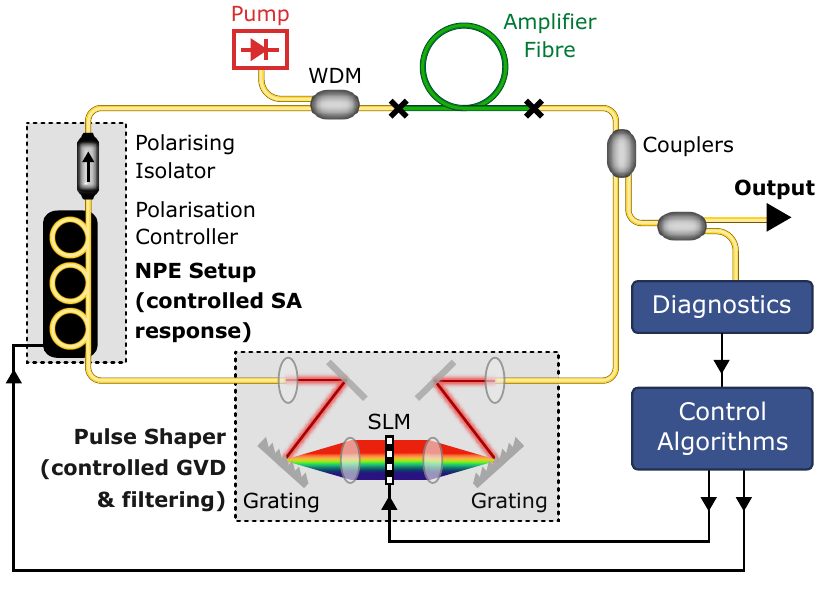}
	\caption{Schematic of a `smart mode-locked fibre laser', showing automated approaches to modify the artificial saturable absorber (SA) response, cavity dispersion and filtering, for self-tuning pulse generation.}
	\label{fig:smart_laser}
\end{figure}

Many of these advances have centred around the inclusion of an intracavity spectral pulse shaper---i.e. a fully-programmable amplitude and/or phase filter, such as a spatial light modulator (SLM) in the Fourier plane of a grating-based dispersive delay line~\cite{Iegorov2016} (Fig.~\ref{fig:smart_laser}); see Ref.~\cite{Boscolo2016} for an excellent recent review.
Phase filtering enables precise control over the cavity dispersion~\cite{Yang2013a} and has led to automated adjustment between soliton, dispersion-managed soliton and dissipative soliton dynamics from a single cavity, with variable pulse durations~\cite{Peng2016a}.
Adaptive intracavity wave-shaping can also enable simpler access to parabolic pulse operating regimes, which are typically more challenging to achieve, e.g. Ref.~\cite{Iegorov2016} located a wave-breaking free operating regime, achieving de-chirped durations of 55~fs, with additional automated optimisation of the pulse quality.
In the absence of a pulse shaper, computer-controlled dispersion/filtering can still be introduced to an extent using automated opto-mechanics (e.g. for grating rotation).

Automation strategies have also been applied to obtain a strong saturable absorber response from NPE~\cite{Radnatarov2013,Olivier2015,Andral2015} and NALM~\cite{Woodward_scirep_2016} designs, by automatically adjusting the waveplate phase-bias to achieve stable mode-locking.
Of greatest promise is the combination of such parameter automation with real-time output monitoring and artificial intelligence concepts, leading to self-tuning lasers that achieve a desired operating state without being explicitly programmed how to achieve this.
This has recently been demonstrated in proof-of-principle experiments using genetic algorithms for mode-locking~\cite{Andral2015, Woodward_scirep_2016} and spectral tuning~\cite{Woodward_ol_2017}.
We expect that further application of intelligent algorithms, e.g. exploiting advances in machine learning~\cite{Brunton2014}, with intracavity amplitude and phase control could lead to a generation of `smart lasers'~\cite{Woodward_scirep_2016} (schematically illustrated in Fig.~\ref{fig:smart_laser}) that may even discover operating regimes that outperform the current state-of-the-art.

\subsection{Opportunities in the Mid-Infrared}
\label{sec:midir}
Until recently, all reported mode-locked fibre lasers operated in the near-infrared (near-IR) below $\sim$2.2~\mum wavelength, limited by the transparency window of silica fibre.
Due to significant demand to push the operating wavelength of fibre pulse sources into the mid-infrared to enable new applications (e.g. in sensing and manufacturing due to the existence of strong mid-IR absorption resonances for many important organic and industrial materials), mid-IR fibre laser development using fluoride (e.g. ZBLAN) fibres is now a growing focus~\cite{Jackson2012}.

This has led to demonstrations of mode-locked fibre lasers operating around 3~\mum using holmium- and erbium-doped ZBLAN fibre~\cite{Frerichs1996, Duval2015, Antipov2016a}.
Despite operating with all-anomalous dispersion, these soliton lasers have generated up to 7.6~nJ femtosecond pulses with 37~kW peak power~\cite{Antipov2016a}; ideal for octave-spanning supercontinuum generation~\cite{Hudson2017} and nonlinear compression to few-cycle pulse durations~\cite{woodward_2017_70fs}.

\begin{figure}[tb]
	\centering
	\sffamily
	\includegraphics{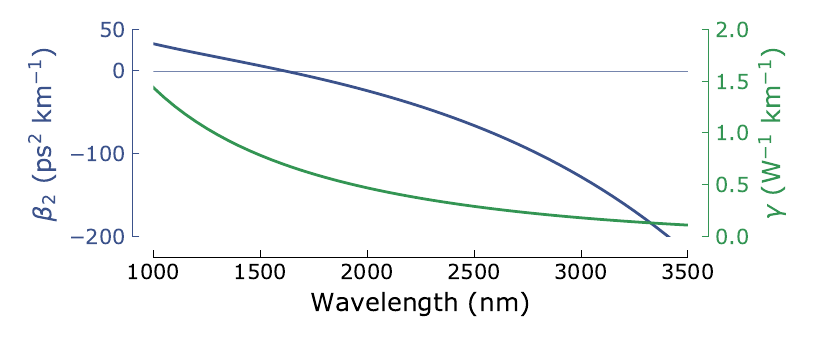}
	\begin{overpic}{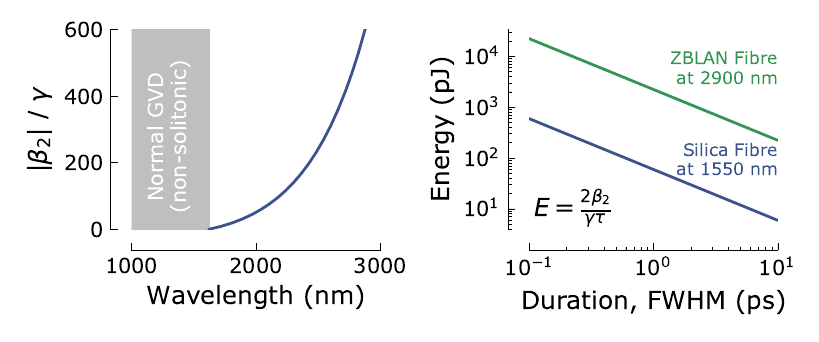}
		\put(0, 80){ {\small (a)} }
		\put(0, 38){ {\small (b)} }
		\put(48, 38){ {\small (c)} }
	\end{overpic}
	\rmfamily
	\caption{Properties of ZBLAN fibre (13~\mum core diameter, 0.13 NA): (a) GVD \& nonlinear parameter; (b) ratio of GVD to nonlinear parameter; (c) soliton energy--duration relationship at 2900~nm, compared to typical single-mode silica fibre (8.2~\mum core diameter, 0.12 NA) at 1550~nm.}
	\label{fig:mid_ir}
\end{figure}

Compared to low-energy near-IR soliton lasers, the greatly improved pulse energies available from mid-IR soliton lasers arise from the increased $\beta_2 / \gamma$ ratio (which limits pulse energy by the area theorem, Eqn.~\ref{eqn:area_theorem}) for typical fluoride fibres at mid-IR wavelengths.
ZBLAN glass is strongly anomalously dispersive in the mid-IR.
In addition, the guided mode area $A_\mathrm{eff}$ is larger at longer wavelengths and ZBLAN possesses a reduced nonlinear index compared to silica ($n_2 =2.1\times10^{-20}$~m$^2$/W compared to 2.7$\times10^{-20}$~m$^2$/W for silica~\cite{Agger2012}), leading to a significant reduction in the nonlinear parameter, $\gamma(\lambda) \propto n_2 / (\lambda  A_\mathrm{eff})$.
Fig.~\ref{fig:mid_ir} shows the dispersion, nonlinearity and $\beta_2 / \gamma$  ratio for a typical single-mode ZBLAN fibre (13~\mum core diameter, 0.13 NA), computed numerically using an eigenmode analysis~\cite{Snyder1983}.
For comparison, a standard single-mode silica fibre at 1550~nm (e.g. $\beta_2=-22$~\pskmN, $\gamma=1.3$~\WkmN) gives $|\beta_2| / \gamma = 17$, whereas the ZBLAN fibre at 2900~nm ($\beta_2=-114$~\pskmN, $\gamma=0.2$~\WkmN) yields $|\beta_2| / \gamma = 570$.
Solitons in mid-IR lasers can thus sustain energies an order of magnitude larger than in near-IR lasers (Fig.~\ref{fig:mid_ir}c).

We expect that even higher energy pulses will be possible by extending the wide range of operating regimes discovered for near-IR lasers (Section~\ref{sec:designs}) to the mid-IR region. 
Dispersion management opportunities are aided by the wide range of dispersive and nonlinear properties available from mid-IR fibres, including anomalous low-nonlinearity fluoride fibres and normal-dispersion high-nonlinearity chalcogenide fibres.
Mid-IR laser transitions also typically offer broader gain bandwidths, which is favourable for ultrashort pulse production and high-energy similariton shaping, although to date, all mid-IR laser gain fibres have been anomalously dispersive, which prohibits parabolic pulse formation.

Finally, we note that while the reported femtosecond mid-IR fibre lasers to date have all employed hybrid bulk-fibre cavities (e.g. with free-space dichroic mirrors, isolators, waveplates etc.), concurrent progress is being made in the development of soft-glass fibre components such as FBGs~\cite{Haboucha2014a, Bharathan2017}, couplers~\cite{Stevens2016, Tavakoli2017} and novel splicing techniques~\cite{Okamoto2011,Thapa2015}, paving the way to all-fibre, ruggedised mode-locked mid-IR sources.

\section{Summary}
In conclusion, we have outlined the various dynamical operating regimes of mode-locked fibre lasers, which can be accessed through interplay between dispersion, nonlinearity and dissipative processes.
The typical characteristics and round-trip pulse evolutions of each regime were reviewed, supported by numerical modelling, and limitations were identified.
Normal dispersion operating regimes were shown to offer significantly improved high-energy performance, overcoming the fundamental restrictions of soliton pulse shaping effects in net-anomalous dispersion lasers.
Both dissipative solitons and similaritons hold promise for next-generation high-energy ultrafast lasers, despite some nonlinear instabilities persisting.
Promising opportunities lie ahead and we expect the performance of mode-locked fibre lasers to continue to improve with additional research into novel nonlinear dynamics, further broadening their applications and importance in scientific, industrial and medical fields.

\section*{Appendix: Numerical Modelling Methods}
\label{sec:appendix}
Simulations in this work are based on a piecewise round-trip model where a discretised complex field envelope $A(\omega)$ on a numerical grid is propagated through models representing each cavity component in turn, with 50\% of the light extracted each round-trip as the output and the remaining light fed back for the next round-trip.
The initial field is propagated over many round trips until convergence to a steady state.
A scalar model is used which assumes a fixed linear polarisation (i.e. using all-polarisation-maintaining fibre), although we note that this approach has also been widely shown to yield good agreement with lasers constructed from non-PM components (where an adjustable polarisation controller is included to compensate for uncontrolled cavity birefringence, discussed in Section.~\ref{sec:pol}).
As it has been shown that the simulated steady-state is influenced by the form of the initial seed field in some cases (e.g. lasers with long cavities~\cite{Yarutkina2013}), here we choose to seed stochastically using a one-photon-per-mode shot noise model, analogous to real-world experimental systems.

Propagation in fibre is described by a generalised nonlinear Schr\"{o}dinger equation (similar to the formalism of Ref.~\cite{Travers2010}), including Raman and shock terms:
\begin{equation}
\label{eqn:gnlse}
\eqalign{
\frac{\partial A(z,\omega)}{\partial z}& = \frac{g(z, \omega)}{2}A(z,\omega)\\
& + i[\frac{\beta_2}{2}(\omega-\omega_0)^2]A(z,\omega)\\
& + i \gamma \left[1 + \frac{\omega - \omega_0}{\omega_0}  \right] {\cal F} \biggl\{ A(z,T)\\
& \times \biggl[  (1-f_r)  |A(z,T)|^2\\
& +  f_r\int_{-\infty}^{+\infty}h_R(T')|A(z,T-T')|^2 dT'  \biggr]   \biggr\}
}
\end{equation}
with fibre GVD $\beta_2$ and nonlinear parameter $\gamma$, and where $h_R$ is the Raman response function~\cite{Hollenbeck2002} with $f_r=0.18$ fractional Raman contribution for silica.
The frequency-domain field $A(\omega)$ and time-domain field $A(t)$ are related by fast Fourier transform ${\cal F}$.
The numerical grid is centred on frequency $\omega_0$ and Eqn.~\ref{eqn:gnlse} is efficiently solved in the Interaction Picture method~\cite{Hult2007} using an explicit Runge-Kutta technique of order 5(4).

Active fibre is simulated by including gain $g(z,\omega)$ in Eqn.~\ref{eqn:gnlse}.
To account for gain saturation, a simple model assuming a saturation energy $E_\mathrm{sat}$ value is used (which is varied to simulate changing the amplifier pump power), with peak gain:
\begin{equation}
g(z) = \frac{g_0}{1 + E(z) / E_\mathrm{sat}}
\end{equation}
where $g_0=30$~dB is the small-signal gain and the pulse energy is $E(z) = \int |A(z,t)|^2 dt$.
Finite gain bandwidth is included by multiplying $g(z)$ with a Lorentzian profile of 50~nm width to form $g(z,\omega)$.
Even though more advanced gain models have been suggested to include the specific spectroscopy of active dopants / the frequency-dependent saturation of ultrashort pulses~\cite{Barnard1994,Paschotta2017}, this method is a reasonable approximation that has been widely shown to yield good agreement with experiments.

The saturable absorber is described by an intensity $I(t)=|A(t)|^2$ dependent transfer function (i.e. acting instantaneously on the pulse) with transmission:
\begin{equation}
\label{eqn:sa}
T(I) = 1 - \frac{\alpha_0}{1 + I / I_\mathrm{sat}}
\end{equation}
where $\alpha_0$ is the modulation depth and $I_\mathrm{sat}$ is the saturation intensity.

\ack
I thank Stuart Jackson, Darren Hudson, Matthew Majewski, Roy Taylor, Edmund Kelleher, Robbie Murray and Tim Runcorn for lively discussions and suggestions on this topic.
Funding through an MQ Research Fellowship is also acknowledged.

\section*{References}

\begin{thebibliography}{100}
	\expandafter\ifx\csname url\endcsname\relax
	\def\url#1{{\tt #1}}\fi
	\expandafter\ifx\csname urlprefix\endcsname\relax\def\urlprefix{URL }\fi
	\providecommand{\eprint}[2][]{\url{#2}}
	
	\bibitem{Okhotnikov}
	Okhotnikov O~G 2012 {\em {Fiber Lasers}\/} (Wiley) ISBN 9783527410392
	
	\bibitem{Fermann2003}
	Fermann M~E, Galvanauskas A and Sucha G 2003 {\em {Ultrafast Lasers: Technology
			and Applications}\/} (Marcel Dekker) ISBN 0824708415
	
	\bibitem{DeSilvestri1984}
	DeSilvestri S, Laporta P and Svelto O 1984 {\em IEEE J. Quantum Electron.\/}
	{\bf QE-20} 533
	
	\bibitem{Kartner1998}
	K{\"{a}}rtner F~X, der Au J~A and Keller U 1998 {\em IEEE J. Sel. Top. Quantum
		Electron.\/} {\bf 4} 159--168
	
	\bibitem{Keller1996}
	Keller U, Weingarten K~J, Kartner F~X, Kopf D, Braun B, Jung I~D, Fluck R,
	Honninger C, Matuschek N and {Aus der Au} J 1996 {\em IEEE J. Quantum
		Electron.\/} {\bf 2} 435--453
	
	\bibitem{Martinez2013a}
	Martinez A and Sun Z 2013 {\em Nature Photon.\/} {\bf 7} 842--845 
	
	\bibitem{Woodward2015_as_2d}
	Woodward R~I and Kelleher E~J~R 2015 {\em Appl. Sci.\/} {\bf 5} 1440--1456
	
	\bibitem{Sobon2015}
	Sobon G 2015 {\em Photon. Res.\/} {\bf 3} A56
	
	\bibitem{Agrawal2013}
	Agrawal G~P 2013 {\em {Nonlinear Fiber Optics}\/} (Academic Press) ISBN
	9780123970237
	
	\bibitem{Dennis1994a}
	Dennis M~L and Duling I~N 1994 {\em Opt. Lett.\/} {\bf 19} 1750--2 
	
	\bibitem{Zakharov1972}
	Zakharov V~E and Shabat A~B 1972 {\em Sov. Phys. JETP\/} {\bf 34} 62
	
	\bibitem{Hasegawa1973}
	Hasegawa A and Tappert F 1973 {\em Appl. Phys. Lett.\/} {\bf 23} 142 
	
	\bibitem{Dudley2001}
	Dudley J~M, Peacock A~C and Millot G 2001 {\em Opt. Commun.\/} {\bf 193}
	253--259
	
	\bibitem{Taylor1992}
	Taylor J~R 1992 {\em {Optical Solitons: Theory and Experiment}\/} (Cambridge
	University Press)
	
	\bibitem{Kelly1991}
	Kelly S~M, Smith K, Blow K~J and Doran N~J 1991 {\em Opt. Lett.\/} {\bf 16}
	1337--9 
	
	\bibitem{Kutz2006}
	Kutz J~N 2006 {\em SIAM Rev.\/} {\bf 48} 629--678 ISSN 0036-1445
	
	\bibitem{Bradley1974}
	Bradley D~J and New G~H~C 1974 {\em Proc. IEEE\/} {\bf 62} 313--345 
	
	\bibitem{Ozgoren2010}
	{\"{O}}zg{\"{o}}ren K and Ilday F~{\"{O}} 2010 {\em Opt. Lett.\/} {\bf 35}
	1296--1298
	
	\bibitem{Horowitz1997}
	Horowitz M, Barad Y and Silberberg Y 1997 {\em Opt. Lett.\/} {\bf 22} 799--801
	
	\bibitem{Jeong2014a}
	Jeong Y, Vazquez-Zuniga L~A, Lee S and Kwon Y 2014 {\em Opt. Fiber Technol.\/}
	{\bf 20} 575--592
	
	\bibitem{Mollenauer1986}
	Mollenauer L~F, Gordon J~P and Islam M~N 1986 {\em IEEE J. Quantum Electron.\/}
	{\bf QE-22} 157--173
	
	\bibitem{Kelly1992}
	Kelly S~M~J 1992 {\em Electron. Lett.\/} {\bf 28} 806--808
	
	\bibitem{Tomlinson1985}
	Tomlinson W~J, Stolen R~H and Johnson A~M 1985 {\em Opt. Lett.\/} {\bf 10}
	457--459
	
	\bibitem{Anderson1992}
	Anderson D, Desaix M, Lisak M and Quiroga-Teixeiro M~L 1992 {\em J. Opt. Soc.
		Am. B\/} {\bf 9} 1358--1361 
	
	\bibitem{Stolen1982b}
	Stolen R~H, Botineau J and Ashkin A 1982 {\em Opt. Lett.\/} {\bf 7} 512--514
	
	\bibitem{Doran1988}
	Doran N~J and Wood D 1988 {\em Opt. Lett.\/} {\bf 13} 56--58 
	
	\bibitem{Fermann1990}
	Fermann M~E, Haberl F, Hofer M and Hochreiter H 1990 {\em Opt. Lett.\/} {\bf
		15} 752--4 
	
	\bibitem{Jiang1999}
	Jiang M, Sucha G, Fermann M~E, Jimenez J, Harter D, Dagenais M, Fox S and Hu Y
	1999 {\em Opt. Lett.\/} {\bf 24} 1074--6
	
	\bibitem{Woodward_prj_2015}
	Woodward R~I, Howe R~C~T, Hu G, Torrisi F, Zhang M, Hasan T and Kelleher E~J~R
	2015 {\em Photon. Res.\/} {\bf 3} A30--A42
	
	\bibitem{Renninger2015}
	Renninger W~H and Wise F~W 2015 {\em IEEE J. Sel. Top. Quantum Electron.\/}
	{\bf 21}
	
	\bibitem{Haus2000}
	Haus H~A 2000 {\em IEEE J. Sel. Top. Quantum Electron.\/} {\bf 6} 1173--1185

	
	\bibitem{Menyuk1987a}
	Menyuk C~R 1987 {\em J. Opt. Soc. Am. B\/} {\bf 12} 614
	
	\bibitem{Collings2000}
	Collings B~C, Cundiff S~T, Akhmediev N~N, Soto-Crespo J~M, Bergman K and Knox
	W~H 2000 {\em J. Opt. Soc. Am. B\/} {\bf 17} 354 
	
	\bibitem{Kobtsev2014b}
	Kobtsev S, Smirnov S, Kukarin S and Turitsyn S 2014 {\em Opt. Fiber Technol.\/}
	{\bf 20} 615--620
	
	\bibitem{Kartner1999}
	K{\"{a}}rtner F~X, Zumb{\"{u}}hl D~M and Matuschek N 1999 {\em Phys. Rev.
		Lett.\/} {\bf 82} 4428
	
	\bibitem{Lecaplain2012}
	Lecaplain C, Grelu P, Soto-Crespo J~M and Akhmediev N 2012 {\em Phys. Rev.
		Lett.\/} {\bf 108} 1--5
	
	\bibitem{Turitsyna2013}
	Turitsyna E~G, Smirnov S~V, Sugavanam S, Tarasov N, Shu X, Babin S~A, Podivilov
	E~V, Churkin D~V, Falkovich G and Turitsyn S~K 2013 {\em Nature Photon.\/}
	{\bf 7} 783--786 
	
	\bibitem{Runge2015}
	Runge A~F~J, Broderick N~G~R and Erkintalo M 2015 {\em Optica\/} {\bf 2} 36
	
	\bibitem{Churkin2015}
	Churkin D~V, Sugavanam S, Tarasov N, Khorev S, Smirnov S~V, Kobtsev S~M and
	Turitsyn S~K 2015 {\em Nat. Commun.\/} {\bf 6} 7004 
	
	\bibitem{Woodward_2016_pre}
	Woodward R~I and Kelleher E~J~R 2016 {\em Phys. Rev. E\/} {\bf 93} 032221
	
	\bibitem{Aguergaray2013a}
	Aguergaray C, Runge A, Erkintalo M and Broderick N~G~R 2013 {\em Opt. Lett.\/}
	{\bf 38} 2644 
	
	\bibitem{Alcock1986}
	Alcock I~P, Ferguson A~I, Hanna D~C and Tropper A~C 1986 {\em Electron.
		Lett.\/} {\bf 22} 268--269 
	
	\bibitem{Geister1988}
	Geister G and Ulrich R 1988 {\em Opt. Commun.\/} {\bf 68} 187--189
	
	\bibitem{Duling1988}
	Duling I~N, Goldberg L and Weller J~F 1988 {\em Electron. Lett.\/} {\bf 24}
	1333--1335
	
	\bibitem{Hanna1989}
	Hanna D, Kazer A, Phillips M, Shepherd D and Suni P 1989 {\em Electron.
		Lett.\/} {\bf 25} 95
	
	\bibitem{Kafka1989}
	Kafka J~D, Baer T and Hall D~W 1989 {\em Opt. Lett.\/} {\bf 14} 1269--1271
	
	\bibitem{Hofer1991}
	Hofer M, Fermann M~E, Haberl F, Ober M~H and Schmidt A~J 1991 {\em Opt.
		Lett.\/} {\bf 16} 502--504 
	
	\bibitem{Duling1991}
	Duling I 1991 {\em Opt. Lett.\/} {\bf 16} 539--541
	
	\bibitem{Duling1994}
	Duling I~N, Chen C~J, Wai P~K~A and Menyuk C~R 1994 {\em IEEE J. Quantum
		Electron.\/} {\bf 30} 194
	
	\bibitem{Matsas1992}
	Matsas V~J, Newson T~P, Richardson D~J and Payne D~N 1992 {\em Electron.
		Lett.\/} {\bf 28} 1391
	
	\bibitem{Noske1992a}
	Noske D~U, Pandit N and Taylor J~R 1992 {\em Electron. Lett.\/} {\bf 28} 2185
	
	\bibitem{Tamura1992}
	Tamura K, Haus H~A and Ippen E~P 1992 {\em Electron. Lett.\/} {\bf 28}
	2226--2228
	
	\bibitem{Kelly1991b}
	Kelly S~M~J, Smith K, Blow K~J and Doran N~J 1991 {\em Opt. Lett.\/} {\bf 16}
	1337--9
	
	\bibitem{Hasegawa1991}
	Hasegawa A and Kodama Y 1991 {\em Phys. Rev. Lett.\/} {\bf 66} 161--164
	
	\bibitem{Okhotnikov2003}
	Okhotnikov O~G, Gomes L, Xiang N, Jouhti T and Grudinin A~B 2003 {\em Opt.
		Lett.\/} {\bf 28} 1522 
	
	\bibitem{Lim2003}
	Lim H, Ilday F~O and Wise F~W 2003 {\em Opt. Lett.\/} {\bf 28} 660--662
	
	\bibitem{Avdokhin2003}
	Avdokhin A~V, Popov S~V and Taylor J~R 2003 {\em Opt. Express\/} {\bf 11}
	265--269
	
	\bibitem{Nelson1995}
	Nelson L~E, Ippen E~P and Haus H~A 1995 {\em Appl. Phys. Lett.\/} {\bf 67}
	19--21 
	
	\bibitem{Sharp1996}
	Sharp R~C, Spock D~E, Pan N and Elliot J 1996 {\em Opt. Lett.\/} {\bf 21}
	881--883
	
	\bibitem{Ober1993}
	Ober M~H, Hofer M and Fermann M~E 1993 {\em Opt. Lett.\/} {\bf 18} 367--369

	
	\bibitem{Tamura1993}
	Tamura K, Ippen E, Haus H and Nelson L 1993 {\em Opt. Lett.\/} {\bf 18}
	1080--1082
	
	\bibitem{Lin1980}
	Lin C, Kogelnik H and Cohen L~G 1980 {\em Opt. Lett.\/} {\bf 5} 476--478
	
	\bibitem{Nelson1997}
	Nelson L~E, Jones D~J, Tamura K, Haus H~A and Ippen E~P 1997 {\em Appl. Phys.
		B\/} {\bf 65} 277--294
	
	\bibitem{Tamura1994}
	Tamura K, Nelson L~E, Haus H~A and Ippen E~P 1994 {\em Appl. Phys. Lett.\/}
	{\bf 64} 149
	
	\bibitem{Namiki1997}
	Yu C~X, Namiki S and Haus H~A 1997 {\em IEEE J. Quantum Electron.\/} {\bf 33}
	649--659 
	
	\bibitem{Nelson1996}
	Nelson L~E, Fleischer S~B, Lenz G and Ippen E~P 1996 {\em Opt. Lett.\/} {\bf
		21} 1759 
	
	\bibitem{Zhang2013l}
	Zhang Z, Cenel C, Hamid R and Ilday F~O 2013 {\em Opt. Lett.\/} {\bf 38}
	956--958 
	
	\bibitem{Nomura2014a}
	Nomura Y and Fuji T 2014 {\em Opt. Express\/} {\bf 22} 12461
	
	\bibitem{Turitsyn2012}
	Turitsyn S~K, Bale B~G and Fedoruk M~P 2012 {\em Phys. Rep.\/} {\bf 521}
	135--203 
	
	\bibitem{Haus1992}
	Haus H~A, Fujimoto J~G and Ippen E~P 1992 {\em IEEE J. Quantum Electron.\/}
	{\bf 28} 2086--2096 
	
	\bibitem{Hofer1992}
	Hofer M, Ober M~H, Haberl F and Fermann M~E 1992 {\em IEEE J. Quantum
		Electron.\/} {\bf 28} 720
	
	\bibitem{Orsila2004}
	Orsila L, Gomes L~A, Xiang N, Jouhti T and Okhotnikov O~G 2004 {\em Appl.
		Opt.\/} {\bf 43} 1902--6 
	
	\bibitem{Chong2006}
	Chong A, Buckley J, Renninger W and Wise F 2006 {\em Opt. Express\/} {\bf 14}
	10095
	
	\bibitem{Zhao2006b}
	Zhao L~M, Tang D~Y and Wu J 2006 {\em Opt. Lett.\/} {\bf 31} 1788--1790 
	
	\bibitem{Zhao2006c}
	Zhao L~M, Tang D~Y, Cheng T~H and Lu C 2006 {\em Opt. Lett.\/} {\bf 31}
	2957--2959
	
	\bibitem{DeMatos2004}
	DeMatos C~J~S, Popov S~V, Rulkov A~B, Taylor J~R, Broeng J, Hansen T~P and
	Gapontsev V~P 2004 {\em Phys. Rev. Lett.\/} {\bf 93} 103901
	
	\bibitem{Chong2008a}
	Chong A, Renninger W~H and Wise F~W 2008 {\em J. Opt. Soc. Am. B\/} {\bf 25}
	140 
	
	\bibitem{Akhmediev2005}
	Akhmediev N and Ankiewicz A 2005 {\em {Dissipative Solitons}\/} (Springer) ISBN
	9783540237822
	
	\bibitem{Grelu2012}
	Grelu P and Akhmediev N 2012 {\em Nature Photon.\/} {\bf 6} 84--92
	
	\bibitem{Renninger2012}
	Renninger W~H, Chong A and Wise F~W 2012 {\em IEEE J. Sel. Top. Quantum
		Electron.\/} {\bf 18} 389
	
	\bibitem{Kieu2009}
	Kieu K, Renninger W~H, Chong A and Wise F~W 2009 {\em Opt. Lett.\/} {\bf 34}
	593
	
	\bibitem{Chong2007}
	Chong A, Renninger W~H and Wise F~W 2007 {\em Opt. Lett.\/} {\bf 32} 2408
	
	\bibitem{An2007}
	An J, Kim D, Dawson J~W, Messerly M~J and Barty C~P~J 2007 {\em Opt. Lett.\/}
	{\bf 32} 2010--2012
	
	\bibitem{Ruehl2008}
	Ruehl A, Kuhn V, Wandt D and Kracht D 2008 {\em Opt. Express\/} {\bf 16}
	3130--3135
	
	\bibitem{Ortac2009}
	Orta{\c{c}} B, Baumgartl M, Limpert J and T{\"{u}}nnermann A 2009 {\em Opt.
		Lett.\/} {\bf 34} 1585 
	
	\bibitem{Tang2015b}
	Tang Y, Chong A and Wise F~W 2015 {\em Opt. Lett.\/} {\bf 40} 2361--2364
	
	\bibitem{Ilday2003b}
	Ilday F~{\"{O}}, Buckley J~R, Lim H, Wise F~W and Clark W~G 2003 {\em Opt.
		Lett.\/} {\bf 28} 1365--1367
	
	\bibitem{Jeong2014b}
	Jeong H, Choi S~Y, Rotermund F, Cha Y~h and Yeom D~i 2014 {\em Opt. Express\/}
	{\bf 22} 22667--22672
	
	\bibitem{Tang2015c}
	Tang M, Wang H, Becheker R, Oudar J, Gaponov D, Godin T and Hideur A 2015 {\em
		Opt. Lett.\/} {\bf 40} 1414--1417
	
	\bibitem{Gaponov2015}
	Gaponov D~A, Dauliat R, Darwich D, Mansuryan T, Jamier R, Grimm S, Schuster K
	and Roy P 2015 {\em J. Opt. Soc. Am. B\/} {\bf 32} 1656
	
	\bibitem{Huang2015b}
	Huang C, Wang C, Shang W, Yang N, Tang Y and Xu J 2015 {\em Sci. Rep.\/} {\bf
		5} 13680
	
	\bibitem{Putnam1998}
	Putnam M~A, Dennis M~L, {Duling Iii} I~N, Askins C~G and Friebele E~J 1998 {\em
		Opt. Lett.\/} {\bf 23} 138--140
	
	\bibitem{Renninger2008a}
	Renninger W~H, Chong A and Wise F~W 2008 {\em Opt. Lett.\/} {\bf 33} 3025--3027
	
	\bibitem{Erkintalo2012}
	Erkintalo M, Aguergaray C, Runge A and Broderick N~G~R 2012 {\em Opt.
		Express\/} {\bf 20} 22669 
	
	\bibitem{Kelleher2009}
	Kelleher E~J~R, Travers J~C, Sun Z, Rozhin A~G, Ferrari A~C, Popov S~V and
	Taylor J~R 2009 {\em Appl. Phys. Lett.\/} {\bf 95} 111108
	
	\bibitem{Tian2009}
	Tian X, Tang M, Shum P~P, Gong Y, Lin C, Fu S and Zhang T 2009 {\em Opt.
		Lett.\/} {\bf 34} 1432
	
	\bibitem{Zhang2009d}
	Zhang M, Chen L~L, Zhou C, Cai Y, Ren L and Zhang Z~G 2009 {\em Laser Phys.
		Lett.\/} {\bf 6} 657--660
	
	\bibitem{Woodward_ptl_2014}
	Woodward R~I, Kelleher E~J~R, Popa D, Hasan T, Bonaccorso F, Ferrari A~C, Popov
	S~V and Taylor J~R 2014 {\em IEEE Photon. Technol. Lett.\/} {\bf 26}
	1672--1675

	
	\bibitem{Woodward_ol_2015_gco}
	Woodward R~I, Kelleher E~J~R, Runcorn T~H, Loranger S, Popa D, Wittwer V~J,
	Ferrari A~C, Popov S~V, Kashyap R and Taylor J~R 2015 {\em Opt. Lett.\/} {\bf
		40} 387--390
	
	\bibitem{Bowen2016b}
	Bowen P, Erkintalo M, Provo R, Harvey J~D and Broderick N~G~R 2016 {\em Opt.
		Lett.\/} {\bf 41} 5270

	
	\bibitem{Anderson1993}
	Anderson D, Desaix M, Karlsson M, Lisak M and Quiroga-Teixeiro M~L 1993 {\em J.
		Opt. Soc. Am. B\/} {\bf 10} 1185--1190
	
	\bibitem{Dudley2007}
	Dudley J~M, Finot C, Richardson D~J and Millot G 2007 {\em Nat. Phys.\/} {\bf
		3} 597--603
	
	\bibitem{Tamura1996}
	Tamura K and Nakazawa M 1996 {\em Opt. Lett.\/} {\bf 21} 68--70
	
	\bibitem{Fermann2000}
	Fermann M~E, Kruglov V~I, Thomsen B~C, Dudley J~M and Harvey J~D 2000 {\em
		Phys. Rev. Lett.\/} {\bf 84} 6010--6013 
	
	\bibitem{Chong2015}
	Chong A, Wright L~G and Wise F~W 2015 {\em Reports Prog. Phys.\/} {\bf 78}
	113901
	
	\bibitem{Oktem2010}
	Oktem B, {\"{U}}lg{\"{u}}d{\"{u}}r C and Ilday F~{\"{O}} 2010 {\em Nature
		Photon.\/} {\bf 4} 307--311 
	
	\bibitem{Renninger2010a}
	Renninger W~H, Chong A and Wise F~W 2010 {\em Phys. Rev. A\/} {\bf 82} 3--6

	
	\bibitem{Bale2010}
	Bale B~G and Wabnitz S 2010 {\em Opt. Lett.\/} {\bf 35} 2466--2468
	
	\bibitem{Aguergaray2010}
	Aguergaray C, M{\'{e}}chin D, Kruglov V and Harvey J~D 2010 {\em Opt.
		Express\/} {\bf 18} 8680--8687
	
	\bibitem{Hirooka2004}
	Hirooka T and Nakazawa M 2004 {\em Opt. Lett.\/} {\bf 29} 498--500 
	
	\bibitem{Finot2007}
	Finot C, Barviau B, Millot G, Guryanov A, Sysoliatin A and Wabnitz S 2007 {\em
		Opt. Express\/} {\bf 15} 15824--15835
	
	\bibitem{Nie2011}
	Nie B, Pestov D, Wise F~W and Dantus M 2011 {\em Opt. Express\/} {\bf 19} 12074

	
	\bibitem{Chu2015}
	Chu K~C, Jiang H~Y and Yang S~D 2015 {\em Opt. Lett.\/} {\bf 40} 5319--5322

	
	\bibitem{Wang2017b}
	Wang Z, Zhan L, Fang X and Luo H 2017 {\em J. Opt. Soc. Am. B\/} {\bf 34} 2325

	
	\bibitem{Ilday2004}
	Ilday F, Buckley J, Clark W and Wise F 2004 {\em Phys. Rev. Lett.\/} {\bf 92}
	213902
	
	\bibitem{Ruehl2005}
	Ruehl A, Hundertmark H, Wandt D, Fallnich C and Kracht D 2005 {\em Opt.
		Express\/} {\bf 13} 6305--6309
	
	\bibitem{KM2017}
	{KMLabs - YFi} \urlprefix\url{https://kmlabs.com/product/y-fi/}
	
	\bibitem{SP2017}
	{Southern Photonics - Fiber Lasers}
	\urlprefix\url{http://www.southernphotonics.com}
	
	\bibitem{Krolopp2016}
	Krolopp A, Csakanyi A, Haluszka D, Csati D, Vass L, Kolonics A, Wikonkall N and
	Szipocs R 2016 {\em Biomed. Opt. Express\/} {\bf 7} 49--53
	
	\bibitem{Bednyakova2013a}
	Bednyakova A~E, Babin S~A, Kharenko D~S, Podivilov E~V, Fedoruk M~P,
	Kalashnikov V~L and Apolonski A 2013 {\em Opt. Express\/} {\bf 21} 20556
	
	\bibitem{Boscolo2012b}
	Boscolo S and Turitsyn S~K 2012 {\em Phys. Rev. A\/} {\bf 043811}
	
	\bibitem{Chang2015a}
	Chang W, Soto-Crespo J~M, Vouzas P and Akhmediev N 2015 {\em J. Opt. Soc. Am.
		B\/} {\bf 32} 1377--1383
	
	\bibitem{Iegorov2016}
	Iegorov R, Teamir T, Makey G and Ilday F~{\"{O}} 2016 {\em Optica\/} {\bf 3}
	1312
	
	\bibitem{Boscolo2016}
	Boscolo S, Peng J and Finot C 2016 {\em Appl. Sci.\/} {\bf 5} 1379
	
	\bibitem{Yang2013a}
	Yang X, Kamal H, Richardson D~J and Petropoulos P 2013 {\em CLEO:2013\/} {\bf
		2} CM1I.1
	
	\bibitem{Peng2016a}
	Peng J and Boscolo S 2016 {\em Proc. SPIE\/} {\bf 9893} 98930I
	
	\bibitem{Radnatarov2013}
	Radnatarov D, Khripunov S, Kobtsev S, Ivanenko A and Kukarin S 2013 {\em Opt.
		Express\/} {\bf 21} 20626
	
	\bibitem{Olivier2015}
	Olivier M, Gagnon M~D and Pich{\'{e}} M 2015 {\em Opt. Express\/} {\bf 23}
	6738--6746
	
	\bibitem{Andral2015}
	Andral U, Fodil R~S, Amrani F, Billard F, Hertz E and Grelu P 2015 {\em
		Optica\/} {\bf 2} 275--278
	
	\bibitem{Woodward_scirep_2016}
	Woodward R~I and Kelleher E~J~R 2016 {\em Sci. Rep.\/} {\bf 6} 37616
	
	\bibitem{Woodward_ol_2017}
	Woodward R~I and Kelleher E~J~R 2017 {\em Opt. Lett.\/} {\bf 42} 2952
	
	\bibitem{Brunton2014}
	Brunton S~L, Fu X and Kutz J~N 2014 {\em IEEE J. Sel. Top. Quantum Electron.\/}
	{\bf 20} 1101408
	
	\bibitem{Jackson2012}
	Jackson S~D 2012 {\em Nature Photon.\/} {\bf 6} 423--431
	
	\bibitem{Frerichs1996}
	Frerichs C and Unrau U~B 1996 {\em Opt. Fiber Technol.\/} {\bf 2} 358--366
	
	\bibitem{Duval2015}
	Duval S, Bernier M, Fortin V, Genest J, Pich{\'{e}} M and Vall{\'{e}}e R 2015
	{\em Optica\/} {\bf 2} 623
	
	\bibitem{Antipov2016a}
	Antipov S, Hudson D~D, Fuerbach A and Jackson S~D 2016 {\em Optica\/} {\bf 3}
	1373--1376
	
	\bibitem{Hudson2017}
	Hudson D~D, Antipov S, Li L, Alamgir I, Hu T, Amraoui M~E, Messaddeq Y,
	Rochette M, Jackson S~D and Fuerbach A 2017 {\em Optica\/} {\bf 4} 1163--1166
	
	\bibitem{woodward_2017_70fs}
	Woodward R~I, Hudson D~D, Fuerbach A and Jackson S~D 2017 {\em Opt. Lett.\/}
	{\bf 42} 4893
	
	\bibitem{Agger2012}
	Agger C, Petersen C, Dupont S, Steffensen H, Lyngs{\o} J~K, {Carsten L
		Thomsen}, Th{\o}gersen J, {Rud Keiding} S and Bang O 2012 {\em J. Opt. Soc.
		Am. B\/} {\bf 29} 635--645
	
	\bibitem{Snyder1983}
	Snyder A~W and Love J 1983 {\em {Optical Waveguide Theory}\/} (Chapman and Hall
	Ltd)
	
	\bibitem{Haboucha2014a}
	Haboucha A, Fortin V, Bernier M, Genest J, Messaddeq Y and Vall{\'{e}}e R 2014
	{\em Opt. Lett.\/} {\bf 39} 3294 
	
	\bibitem{Bharathan2017}
	Bharathan G, Woodward R~I, Ams M, Hudson D~D, Jackson S~D and Fuerbach A 2017
	{\em Opt. Express\/} {\bf 25} 30013
	
	\bibitem{Stevens2016}
	Stevens G and Woodbridge T 2016 {\em Proc. SPIE\/} {\bf 9730} 973007
	
	\bibitem{Tavakoli2017}
	Tavakoli F, Rekik A, Rochette M and Member S 2017 {\em IEEE Photon. Technol.
		Lett.\/} {\bf 29} 735--738
	
	\bibitem{Okamoto2011}
	Okamoto H, Kasuga K and Kubota Y 2011 {\em Opt. Lett.\/} {\bf 36} 1470--1472
	
	\bibitem{Thapa2015}
	Thapa R, Gattass R~R, Nguyen V, Chin G, Gibson D, Kim W, Shaw L~B and Sanghera
	J~S 2015 {\em Opt. Lett.\/} {\bf 40} 5074
	
	\bibitem{Yarutkina2013}
	Yarutkina I~A, Shtyrina O~V, Fedoruk M~P and Turitsyn S~K 2013 {\em Opt.
		Express\/} {\bf 21} 12942 
	
	\bibitem{Travers2010}
	Travers J~C, Frosz M~H and Dudley J~M 2010 {Nonlinear fibre optics overview}
	{\em Supercontinuum Generation in Optical Fibers\/} ed Dudley J~M and Taylor
	J~R (Cambridge)
	
	\bibitem{Hollenbeck2002}
	Hollenbeck D and Cantrell C~D 2002 {\em J. Opt. Soc. Am. B\/} {\bf 19} 2886

	
	\bibitem{Hult2007}
	Hult J 2007 {\em J. Light. Technol.\/} {\bf 25} 3770--3775
	
	\bibitem{Barnard1994}
	Barnard C, Myslinski P, Chrostowski J and Kavehrad M 1994 {\em IEEE J. Quantum
		Electron.\/} {\bf 30} 1817--1830
	
	\bibitem{Paschotta2017}
	Paschotta R 2017 {\em Opt. Express\/} {\bf 25} 19112--19116
	
\end{thebibliography}

\providecommand{\newblock}{}

\end{document}